\documentclass[aps,pra,10pt,twocolumn,floatfix,nofootinbib,
groupedaddress,longbibliography,nobibnotes]{revtex4-2}

\usepackage{amsmath}
\usepackage{mathtools}
\usepackage{wasysym}
\usepackage{newtxmath}
\usepackage{mathrsfs}

\usepackage{array} 
\usepackage[T1]{fontenc}
\usepackage{newtxtext}
\usepackage{dsfont}                 
\usepackage{bbold}                  
\usepackage[normalem]{ulem}         

\usepackage{enumerate}
\usepackage[shortlabels]{enumitem}

\usepackage[dvipsnames,table]{xcolor}
\usepackage{graphicx}
\graphicspath{{figs/}}              

\usepackage[flushleft]{threeparttable}
\usepackage{multirow}

\usepackage{physics}                
\usepackage{csquotes}               
\usepackage{comment}

\usepackage[caption=false]{subfig}
\usepackage[colorlinks=true,citecolor=cyan,linkcolor=magenta,filecolor=magenta]{hyperref}
\usepackage[capitalize]{cleveref}

\usepackage{tikz}
\usetikzlibrary{decorations.pathmorphing}
\usetikzlibrary{arrows,decorations.markings,cd}

\usepackage{orcidlink}

\DeclareMathAlphabet{\mathbbold}{U}{bbold}{m}{n}

\newcommand{\be}{\begin{equation}}
\newcommand{\ee}{\end{equation}}

	\def\ket#1{|{#1}\rangle}		

\usepackage{bm}

\definecolor{TB}{rgb}{0.93,0.47,0.2}

\begin{document}




\title{Superconductivity in hyperbolic spaces:\texorpdfstring{\vspace{1mm}\\}{} Cayley trees, hyperbolic continuum, and BCS theory}

\author{Mykhailo Pavliuk}
\email{mykhailo.pavliuk@uzh.ch}
\author{Tom\'{a}\v{s} Bzdu\v{s}ek\,\orcidlink{0000-0001-6904-5264}}
\email{tomas.bzdusek@uzh.ch}
\author{Askar Iliasov}
\email{askar.iliasov@uzh.ch}
\affiliation{Department of Physics, University of Zurich, Winterthurerstrasse 190, 8057 Zurich, Switzerland}


\begin{abstract}


We investigate $s$-wave superconductivity in negatively curved geometries, focusing on Cayley trees and the hyperbolic plane. 
Using a self-consistent Bogoliubov--de Gennes approach for trees and a BCS treatment of the hyperbolic continuum, we establish a unified mean-field framework that captures the role of boundaries in hyperbolic spaces. 
For finite Cayley trees with open boundaries, the superconducting order parameter localizes at the edge while the interior can remain normal, leading to two distinct critical temperatures: $T_\textrm{c}^\textrm{edge} > T_\textrm{c}^\textrm{bulk}$. 
A corresponding boundary-dominated phase also emerges in hyperbolic annuli and horodisc regions, where radial variations of the local density of states enhance edge pairing. 
We also demonstrate that the enhancement of the density of states at the boundary is significantly more pronounced for the discrete tree geometry.
Our results show that, owing to the macroscopic extent of the boundary, negative curvature can stabilize boundary superconductivity as a phase that persists in the thermodynamic limit on par with the bulk superconductivity.
These results highlight fundamental differences between bulk and boundary ordering in hyperbolic matter, and provide a theoretical framework for future studies of correlated phases in negatively curved~systems.
\end{abstract}

\maketitle


\section{Introduction}

Lattice geometry plays a crucial role in determining the electronic band structure of quantum systems, which in turn is the starting point for treating the effect of interactions between the particles.
Recent experimental works with circuit quantum electrodynamics~\cite{Kollar:2019}, electric-circuit networks~\cite{Lenggenhager:2021,Zhang:2022,Zhang:2023,Chen:2023b,Chen:2023c}, planar microwave waveguides~\cite{Chen:2024}, and silicon photonics~\cite{Huang:2024} made it possible to also realize two-dimensional (2D) hyperbolic lattices, which constitute a special geometry with an emergent negative curvature~\cite{Boettcher:2020}.
A regular hyperbolic lattice~\cite{Schrauth:2024} is specified by its Schl\"{a}fli symbol $\{p,q\}$, which means that a number $q$ of regular $p$-sided polygons meet at each vertex~\cite{Coxeter:1957,Boettcher:2022},  with the two integers conditioned by $(p-2)(q-2)>4$.
Hyperbolic lattices are unique by combining a high amount of crystalline symmetry with non-commutativity of translations, which significantly increases the complexity of capturing their quantum mechanics~\cite{Maciejko:2021}, leaving one to wonder about the interplay of hyperbolicity with band theory and correlations.

At the single-particle level, the extension of Bloch's theorem to hyperbolic lattices is in principle mathematically understood~\cite{Maciejko:2022,Kienzle:2022,Nagy:2024}; nevertheless, advanced strategies turned to be necessary to adequately compute the bulk spectra of hyperbolic Hamiltonians, including trace formulas~\cite{Mnev:2006,Boettcher:2022b}, converging periodic boundary conditions~\cite{Lux:2022,Lux:2023}, continued fraction expansions~\cite{Mosseri:2023}, and coherent sequences of translation subgroups~\cite{Lenggenhager:2023,HyperCells,HyperBloch}.
In addition, the negative curvature implies that the boundary of any finite hyperbolic system constitutes a macroscopic fraction of its volume; therefore, the choice of open vs.~periodic boundary condition can drastically alter the nature of the spectrum.
Much activity in this direction was focused towards the nature of topological~\cite{Yu:2020,Urwyler:2022,Liu:2022,Chen:2023,Tummuru:2024,Liu:2023, Tao:2023} and flat~\cite{Kollar:2020,Bzdusek:2022,Mosseri:2022,He:2024,Yuan:2024,Guan:2025} energy bands. 
Further single-particle aspects, including orbital coupling to magnetic field~\cite{Ikeda:2021,Stegmaier:2022}, localization due to disorder~\cite{Curtis:2025,Li:2024,Chen:2024b}, topological linear response~\cite{Sun:2024}, entanglement entropy~\cite{Huang:2025}, non-Hermitian skin effects~\cite{Sun:2023,Shen:2025}, have also been extended to hyperbolic lattices, and a connection between hyperbolic lattices and the Yang-Mills theory was pointed out~\cite{Shankar:2024}.

In contrast, much less is presently known about the interplay of negative curvature with many-body correlations.
Hyperbolic analogs of paradigm magnetic Hamiltonians, including the quantum Ising, XY, and Heisenberg models, have been investigated with mean-field theory, spin-wave quantum, and quantum Monte Carlo~\cite{Daniska:2016,Daniska:2018,Gotz:2024}, although one should bear in mind that their validity remains unsettled, since the strong dependence on boundary conditions in hyperbolic lattices complicates the interpretation of such results.
In addition, the Kramers-Wannier duality breaks down for Ising models in hyperbolic lattices with open boundary condition~\cite{Placke:2023}, which relates to the appearance of an additional boundary-sensitive phase at the intermediate regime between the conventional high-temperature (paramagnetic) and low-temperature (ferromagnetic) phase~\cite{Wang:2025}.
Hyperbolic variants of the Fermi- and Bose-Hubbard models have likewise been explored~\cite{Gluscevich:2023,Roy:2024,Zhu:2021,Dutkiewicz:2025}, although their interpretation should be viewed with care given the early stage of this line of research.
Some of these concerns are overcome in the studies of exactly solvable models, including hyperbolic surface codes~\cite{Delfosse:2013,Breuckmann:2016,Breuckmann:2017,Higgott:2024} and hyperbolic Kitaev models~\cite{Lenggenhager:2025,Dusel:2025,Mosseri:2025,Vidal:2025}.
In addition, correspondence between quantum mechanics
of hyperbolic lattices and one-dimensional interacting aperiodic chains at the holographic boundary of hyperbolic lattices has also been investigated~\cite{Basteiro:2022,Basteiro:2023,Flicker:2020,Asaduzzaman:2020}. 

In this work, and in the associated Ref.~\citenum{Bashmakov:2025}, we extend the research of correlations in hyperbolic spaces by investigating the formation of the superconducting condensate.
Our focus lies entirely with $s$-wave superconductivity arising from an attractive on-site (or point-contact) Hubbard interaction between spin-$\tfrac{1}{2}$ particles.
Motivated by the prediction of boundary superconductivity in Euclidean crystals~\cite{Babaev:2020,Barkman:2022, Croitoru:2020, Talkachov:2023,Hainzl:2022}, 
and considering the macroscopic nature of the boundary of hyperbolic lattices, we are specifically interested in the possibility of boundary superconductivity in hyperbolic spaces. 
The anticipation that there could be an intermediate regime where the superconducting condensate forms only near the boundary but not in the bulk is further amplified by the existence of the intermediate phase in the hyperbolic Ising model~\cite{Wang:2025}, which under open boundary conditions with no bias results in magnetic ordering only near the boundary. To investigate the problem, we adopt a mean-field approach to describing superconductivity, such as the Bogolyubov-de Gennes (BdG) or the Bardeen, Cooper, and Schrieffer (BCS) theories.

 \begin{figure}[t]
    \centering
    \includegraphics[width=\linewidth]{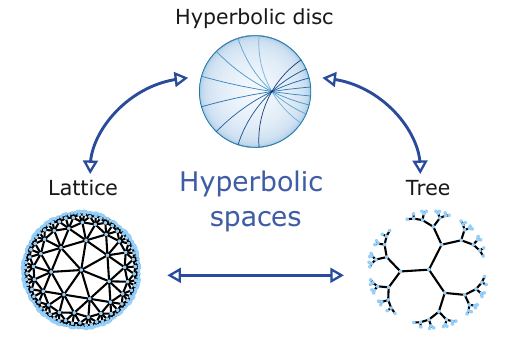}
    \caption{Possible hyperbolic spaces. In this paper, we focus on trees and continuous hyperbolic spaces.
    }
\label{fig:title_figure}
\end{figure}

In the present manuscript, we pay special attention to those realizations of the hyperbolic space where the BCS problem can be treated analytically or semi-analytically (three
possible choices of hyperbolic spaces are shown in Fig.~\ref{fig:title_figure}).
On one hand, we present a
formulation of an exact self-consistent equation for the energy gap in terms of the density of states, which applies for any vertex-transitive graphs, including among others
hyperbolic lattices with periodic boundary conditions.
On the other hand, to treat analytically the case of open boundary conditions, we consider two geometries. 
First, we consider the limit $p \to \infty$ of the hyperbolic $\{p,q\}$ lattice, which corresponds to an infinitely branching tree graph known as the Bethe lattice.
The version of a Bethe lattice that terminates at a finite graph distance from a selected root node (corresponds to a `disk' of the Bethe lattice with open boundary condition) is commonly called the Cayley tree. 
Second, we assume electrons moving in the hyperbolic continuum, i.e., in the absence of a specific hyperbolic lattice, where the negative curvature is implemented through the metric tensor on the two-dimensional manifold.
For both specified models of a hyperbolic space with open boundary, we indeed identify the intermediate phase where the superconducting condensate is localized only at the boundary. However, we should underline that the discrete tree geometry provides more pronounced enhancement of local density of states on the boundary, and hence is more susceptible to the existence of a separate boundary phase.
The results presented here are complemented by Ref.~\citenum{Bashmakov:2025}, which investigates $s$-wave superconductivity of the attractive Hubbard model on hyperbolic $\{p,q\}$ lattices with open boundary conditions and that further employs the Ginzburg-Landau theory to describe the possible behavior of the superconducting condensate near the boundary of a hyperbolic space irrespective of the microscopic details.

The manuscript is structured as follows. 
We divide the article into two main parts: the study of Cayley trees via BdG theory \ref{sec:lattices} and the study of continuous spaces via BCS theory \ref{sec:cont_spaces}. 
For each choice of geometry, we first consider the case of uniform spaces, i.e.~in the absence of a boundary, and show that the mean-field approach reproduces the BCS gap equation (Secs.~\ref{sec:uni_lat} and~Sec.~\ref{sec:uni_space}) with the curvature of the underlying space entering only through the density of states. 
Second, for both the discrete and the continuous case, we focus on open boundary conditions (Sec.~\ref{sec:cayley} and Sec.~\ref{sec:cont_spaces}).
In our investigation of Cayley trees, we introduce a symmetry-adapted block decomposition that makes self-consistent BdG calculations feasible at large radial sizes. 
This technique allows us to demonstrate the existence of two distinct critical temperatures for the bulk and the boundary, where the ratio of boundary and bulk critical temperatures is significantly higher than reported in flat (i.e., Euclidean) systems.
In the continuum case, we perform exact calculations of the local density of states in the horodisc region and find that the boundary enhancement is controlled by the curvature of the hyperbolic space. 
Small curvature leads to boundary enhancement analogous to that in the flat 2D systems, while the extremely curved limit corresponds to stronger boundary amplification analogous to that in the flat 1D case. 
Finally, we numerically solve the BCS equations to find robust boundary-localized superconductivity persisting for temperatures higher than the bulk critical temperature.

\section{Tree graphs}\label{sec:lattices}

Tree graphs provide the simplest discrete setting in which negative curvature manifests itself. 
In this section, we use trees to develop a picture of how hyperbolic geometry can affect superconductivity at the mean-field level. 
To that end, we consider in Sec.~\ref{sec:uni_lat} a general uniform lattice, of which the infinite regular trees---known as Bethe lattices---are an example. 
In this case, the gap equation follows the usual BCS theory for flat spaces and can be expressed in terms of the known single-particle density of states. 
Subsequently, in Sec.~\ref{sec:cayley} we turn to finite Cayley trees with open boundaries. 
Therein, after first introducing analytic tools for capturing the spectra of such finite trees, we explicitly showcase the appearance of boundary-only superconductivity within an extended range of temperatures.

\subsection{Bethe lattices}\label{sec:uni_lat}

We begin with uniform lattices, focusing on the Bethe lattice (the infinite regular tree) as the case of primary interest.
Mean-field superconductivity on general graphs is introduced within the BdG framework.
We show that for a uniform lattice, the gap equation depends only on the single-particle density of states, and we provide the solution for a Bethe lattice.

The starting point of our study of the $s$-wave superconductivity is the Hubbard model with on-site attractive interactions:
\begin{equation}\label{eq:Hubbard_model}
    H=t\sum_{\langle i,j \rangle\sigma} c^{\dagger}_{i\sigma}c_{j\sigma}-\mu\sum_{i\sigma}n_{i\sigma}+U\sum_{i} n_{i\uparrow} n_{i\downarrow}
\end{equation}
where $c^\dagger_{i\sigma},\,c_{i\sigma}$ are the creation and the annihilation operator of a fermion with spin $\sigma=\uparrow,\downarrow$ at site $i$, $n_{i\sigma}=c^{\dagger}_{i\sigma} c_{i\sigma}$ is the particle number operator, $t=-1$ is the nearest-neighbor hopping, $\mu$ is the chemical potential, and $U>0$ is the strength of attractive on-site interactions. 

We investigate the Hubbard model within the mean-field approximation employing the BdG 
approach~\cite{BdG_book}. 
After performing the Bogolyubov transformation and introducing local pairing amplitude $\Delta_i$, the eigenvalue equation for the
BdG Hamiltonian can be written in the form:
\begin{gather}
    H_\textrm{BdG}\begin{pmatrix}
    u_{n}\\v_{n}
    \end{pmatrix}=\begin{pmatrix}
    h-\mu&\mathbb{\Delta}\\ \mathbb{\Delta}&-h+\mu
    \end{pmatrix}    \begin{pmatrix}
    u_{n}\\v_{n}
    \end{pmatrix}=E_{n}\begin{pmatrix}
    u_{n}\\v_{n}
    \end{pmatrix},
\label{eq:BdG_Ham}
\end{gather}
where the matrix $h$ represents the free part of the Hamiltonian [i.e, the first sum in Eq.~(\ref{eq:Hubbard_model}) with omitted spin indices],
$\mathbb{\Delta}=\mbox{diag}(\Delta_1,\dots\Delta_\mathcal{N})$ is a diagonal matrix consisting of values of the order parameter, and $\mathcal{N}$ is the total number of lattice sites. 
The diagonal structure of $\delta$ is the consequence of the on-site Hubbard interaction~\cite{BdG_book}.
We assume the absence of
spin-orbit coupling and pairing into the singlet state, allowing us to omit the spin indices. 
The corresponding self-consistent equations for the superconducting order parameter $\Delta$ are as follows \cite{BdG_book}:
\begin{gather}
    \Delta_i=\frac{U}{2}\sum_{n}u_{n,i} v^{*}_{n,i} \tanh(\frac{E_n}{2 T}), \label{eq:selfcons_delta}
\end{gather}
where the sum is over all eigenstates of $H_\textrm{BdG}$.

We next narrow our attention to so-called uniform lattices, i.e., lattices in which
any two vertices $v_1$ and $v_2$ are related by a symmetry.
In graph theory, such systems are also described as vertex-transitive and the symmetry is usually called an automorphism.
In uniform lattices, the solution of the gap equation should be constant, i.e., $\Delta_i \equiv \Delta$, since symmetry allows us to relate the gap amplitude $\Delta_i$ at any two sites $i$, and we anticipate the $s$-wave superconducting state to transform in the trivial representation of the symmetry group.
In this case, we can write the gap equation in closed integral form (see Appendix \ref{app:uniform_lattices} for derivation):
\be
\label{eq:gapselfconst_uniform}
\Delta=U\!\!\int^{\infty}_{-\infty}\!\frac{\Delta\nu(\lambda)}{2\sqrt{(\lambda-\mu)^2+\Delta^2}}\tanh\Big(\frac{\sqrt{(\lambda-\mu)^2+\Delta^2}}{2 T}\Big)d\lambda,
\ee
where $\nu(\lambda)$ is the density of states of the lattice. One can see that Eq.~\eqref{eq:gapselfconst_uniform} is the gap equation of the BCS theory \cite{Leggett:2006,Noda:2015}. From this observation, we readily conclude that the critical temperature in the weak coupling regime [$U\nu(\mu)\ll 1$] can be captured
by the BCS expression:
\begin{equation} 
\label{eqn:BCS-TC}
T_{c}\sim e^{-\frac{1}{U\nu(\mu)}}.
\end{equation}
However, since the gap equation~\eqref{eq:gapselfconst_uniform} is derived only for vertex-transitive graphs, it correctly describes only bulk superconductivity, and it cannot be used for studying superconductivity near the boundaries.

 \begin{figure}
    \centering
    \hspace*{-0.5cm}
    \includegraphics[width=1.1\linewidth]{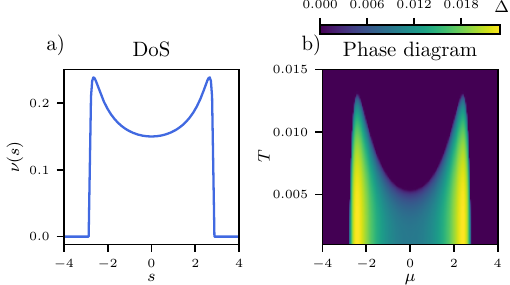}       
    \caption{Density of states (left panel) and phase diagram (right 
    panel) for Bethe lattice with degree of vertices $q=K+1=3$ and for Hubbard potential $U=1$.}
    \label{fig:bethe_example}
\end{figure}

In the
particular case of the Bethe lattice with connectivity $K\ge2$ (defined such that $K+1$ is the degree $q$ of the vertices), we can express the density of states explicitly~\cite{Mnev:2006,Mahan-2001-PRB}:  
\be
\nu_{K}(\lambda)=\frac{1}{2\pi}\frac{\sqrt{4K-\lambda^2}}{(K+1)^2-\lambda^2}.
\ee
Therefore, for the Bethe lattice, the gap equation takes the specific form:
\be
\Delta\!=\!\frac{U}{4\pi}\!\int^{2\sqrt{K}}_{-2\sqrt{K}}\!\frac{\Delta\sqrt{4K-\lambda^2}}{[(K{+}1)^2{-}\lambda^2]\xi(\lambda,\Delta)}\tanh\!\Big(\frac{\xi(\lambda,\Delta)}{2 T}\Big)d\lambda,
\ee
where we introduced a new variable $\xi$, defined as $\xi(\lambda,\Delta) = \sqrt{(\lambda-\mu)^2 + \Delta^2}$.
The density of states and the phase diagram for the Bethe lattice with connectivity $K=2$
are shown in Fig.~\ref{fig:bethe_example}. 
The self-consistent equations were solved for $U=1$.

\subsection{Cayley trees}\label{sec:cayley}

When considering the effect of boundary on the superconductivity in Cayley trees, we cannot use the assumption of uniformity leading to Eq.~\eqref{eq:gapselfconst_uniform}. Nevertheless, a significant simplification of the gap equation can be achieved by decomposing the nearest-neighbor Hamiltonian defined on the Cayley tree into a block diagonal form. 
Such a decomposition is accomplished by applying the construction of symmetric and non-symmetric eigenstates on a Cayley tree as introduced in Refs.~\citenum{Mahan-2001-PRB,Aryal-IOP-2020,Hamanaka:2024}, and also developed and applied in other works \cite{Aizenman:2006,Petrova:2016,Weststrom:2023}. 
This approach allows us to avoid the exact diagonalization of the full BdG Hamiltonian, and we are able to numerically solve the self-consistent equations for large trees containing ${\sim}10^{100}$ sites.

Our treatment of finite trees with open boundary condition is structured as follows.
We begin in Sec.~\ref{sec:sym-basis} with introducing the symmetry adapted basis of states, which enables an analytic treatment of the single-particle spectra on the Cayley tree. 
In Sec.~\ref{sec:BdG-on-Cayley} we utilize the symmetry-adapted basis to rewrite the BdG equation on Cayley trees into a form which is easily numerically tractable for systems with hundreds of layers.
As the next step, we include in Sec.~\ref{sec:Cayley-results} a discussion of the numerical results, which notably include an extended temperature range associated with a boundary-only superconducting~order. Finally, in Sec.~\ref{sec:Cayley_ldos} we discuss the connection of our findings with the structure of local density of states on Cayley trees.

\subsubsection{Symmetric and nonsymmetric states}
\label{sec:sym-basis}

To set up the stage for solving the Hubbard model on Cayley trees, we start with introducing a symmetry-adapted basis of states that block-diagonalize the nearest-neighbor tight-binding Hamiltonian $h$ on the tree. Our treatment follows closely the exposition in Ref.~\citenum{Hamanaka:2024}. The nearest-neighbor Hamiltonian reads:
\be
\label{eq:H_on_trees_posbasis}
h=-\sum_{\langle i,j \rangle} c^{\dagger}_i c_j,
\ee
 where we have set the nearest-neighbor hopping amplitude to $-1$, indices $i,j$ denote nodes of the tree, and the notation $\langle i,j \rangle$ denotes a pair $(i, j)$ connected by an edge. 
 In the following discussion, we use the words `node' and `site' interchangeably on tree graphs.

\begin{figure}[t]
    \centering
    \includegraphics[width=\linewidth]{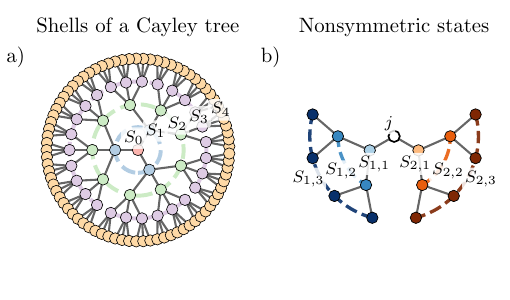}
    \vspace{-1cm}
    \caption{(a) Definition of shells of a Cayley tree. 
    These shells are also used directly in the construction of the symmetric basis states in  Eq.~\eqref{eq: symm basis}. 
    The number of shells in the example is set to $M=4$. 
    (b) Definition  of shells used for constructing nonsymmetric basis states emanating from the seed node $\ket{j}$ as in Eq.~\eqref{eq: nonsymm basis}.
    }
\label{fig:cayley_states}
\end{figure}

A convenient way to introduce the (non)symmetric states is to divide the Cayley tree into radial shells $S_l$ as illustrated in Fig.~\ref{fig:cayley_states}(a). Each shell is defined as the set of nodes with the same distance $l$ from the root of the tree. Then the Hamiltonian~\eqref{eq:H_on_trees_posbasis} can be written using 
operators propagating the shell $l$ to the shells $l+1$ and $l-1$:
\be
H=P_0+\sum^{M-1}_{l=1} (P^{\dagger}_l+P_l)+P^{\dagger}_M.
\ee
where $M$ is the total number of shells, and we choose the convention that the center of the tree has index $0$. The operators $P_l$ propagate the wave functions localized on the $l$-th shell forward from the center, and the operators $P^{\dagger}_l$ propagate the wave functions backward to the center. The forward operators $P_l$ are defined as follows:
\begin{gather}
\label{eq:forward_op}
P_l=\sum_{\substack{\langle i,j \rangle \,\textrm{such that:} \\ i\in S_l \;\textrm{and}\; j\in S_{l+1}}} c^{\dagger}_j c_i,
\end{gather}
where the sum is taken over nearest-neighbor
nodes lying in consecutive shells $l$ and $l+1$. The backward operators $P^{\dagger}_l$ are the Hermitian conjugates of $P_l$ by construction. 
Let us remark
that the operators $P_l$ and $P^{\dagger}_l$ themselves are non-Hermitian and have nontrivial action only on the wave functions that have non-vanishing support on shell $S_l$.

Now, we are ready to construct the (non)symmetric basis states. 
The general procedure in the construction is to choose an initial normalized state $\phi_0$ localized at the $l$-th shell $S_l$ and to propagate it forward:
\be\label{eq:propagation_forward}
\phi'_n=\Big[P_{l+n} \cdot\ldots\cdot P_{l+1} P_l\Big]\phi_0.
\ee
The obtained states $\phi'_n$ are, in general, not normalized. 
After normalization, one obtains the state:
\be\label{eq:states_normed}
\phi_n=\frac{\phi'_n}{||\phi'_n||}.
\ee
Particular choices of $\phi_0$ generate the set $\{\phi_0,\phi_1,\ldots,\phi_{M+1-l}\}$  that forms a sector of (non)symmetric states. 
It is worth noting that the construction of (non)symmetric states described in this section coincides with the Lanczos diagonalization algorithm \cite{Koch:2019} for the considered choice of initial vectors~$\phi_0$.

We start by building the set of symmetric basis states. The initial state $\phi_0$ for symmetric basis states is the position-basis state localized at the root of the tree (central node):
\begin{equation}
    \phi_0= \ket{0}.
\end{equation}
Since it is also the initial symmetric state, we introduce the following notation:
\be
|0)\coloneqq \ket{0}.
\ee
Throughout the discussion of tree graphs, 
we use $\ket{\cdots}$ and $|\cdots)$ to denote the position basis states resp.~the symmetry-adapted (i.e., symmetric and nonsymmetric)
basis states. 
Applying the procedure given by Eq.~\eqref{eq:propagation_forward} to the state $\phi_0=|0)$, we obtain the complete set of symmetric basis states, given by the exact expreesions:
\begin{align}
\label{eq: symm basis}
    | l) \coloneqq \frac{1}{\sqrt{K^{l-1}}} \sum_{i\in S_l} |i\rangle
\end{align}
The symmetric states $|l)$ form $M\,{+}\,1$  orthonormal~states.

Next, we generate the remaining basis states, which we call nonsymmetric basis states. 
We construct an initial nonsymmetric state $\phi_0$ in the following way.
We choose a node $j$ that does not lie 
in the last ($M$-th) shell as the `seed', and we consider the $K$ branches (or the $K+1$ branches if $j$ is the central node) emanating from this node. The key idea for constructing the nonsymmetric states is to weight the branches emanating outward 
from the node $j$ in such a way that the action of the Hamiltonian~(\ref{eq:H_on_trees_posbasis}) results in a destructive interference of the hopping processes at $j$. 

Specifically, if $j$ lies in the $l_{j}$-th shell, then we can choose $\phi_0$ localized at the nodes in the $(l_{j}\,{+}\,1)$-th shell that are direct descendants of node $j$. Then, in the language of the $P$ operators, the condition of destructive interference means~that
\be\label{eq:P_on_nonsym_states}
P^\dagger_{l_{j}+1} \phi_0 =0.
\ee 
The space of possible $\phi_0$ has dimension $K$ if $j$ is the central node $|0\rangle$, and dimension $(K\,{-}\,1)$ if $j$ is not the central node. 
We introduce an orthonormal 
basis in this space, and we call the resulting
basis states as the first nonsymmetric states associated with the node $j$. 
To achieve the orthonormalization,
we choose a $K$-th root of unity $\varpi$ for any $j$ except the central node or [$(K\,{+}\,1)$-th root of unity $\varpi$ if $j$ is the center of the tree]. 
This allows us to express the first nonsymmetric states associated with the site $j$ as follows:
\begin{equation}
\label{eq:nonsymstate_betaneq0}
\!|1,\varpi)_j \!=\! 
\left\{\begin{array}{l}
\!\!\frac{1}{\sqrt{K+1}}\!\sum^{K+1}_{m=1}\! \varpi^m|\gamma_m\rangle \; \textrm{with $\varpi = e^{\frac{2\pi i k}{K+1}}$} \, \textrm{for $j = 0$\!} \\ 
\!\!\frac{1}{\sqrt{K}}\!\sum^{K}_{m=1}\! \varpi^m|\gamma_m\rangle \;\textrm{with $\varpi = e^{\frac{2\pi i k}{K}}$} 
\, \textrm{for $j \neq 0$,}
\end{array}\right.
\end{equation}
where $k$ is a positive integer.
We have further used $\gamma_m$ to label nodes in the $(l_{j}\,{+}\,1)$-th shell which are direct descendants of the node $j$, and 
index $m$ counts the branches emanating from $j$ and takes values $m\in\{1,\ldots, K\}$ if $j\neq 0$ and $m\in\{1,\ldots, K+1\}$ if $j = 0$. One can verify that any root of unity except the trivial root $\varpi = 1$ satisfies
the condition~\eqref{eq:P_on_nonsym_states}.

The remaining nonsymmetric states are obtained by applying Eqs.~\eqref{eq:propagation_forward} and~\eqref{eq:states_normed}, and can be written in closed form. 
If $j$ is the center of the tree, the nonsymmetric states are:
\begin{align}\label{eq: nonsymm basis l=0}
    |r,\varpi)_0 \coloneqq \frac{1}{\sqrt{(K+1)K^{r-1}}} \sum^{K+1}_{m=1}\varpi^m\sum_{i\in S_{m,r}}|i\rangle,
\end{align}
where by $S_{m,r}$ we denote the $r$-th shell of the branch $m$ starting from the node $j$. The index $r$ takes values in $\{1,\ldots,M\}$. These states form a set of $KM$ nonsymmetric orthonormal states associated with the center node.

The nonsymmetric states corresponding to any node except the center are:
\begin{align}\label{eq: nonsymm basis}
    |r,\varpi)_{j\neq 0} \coloneqq \frac{1}{\sqrt{K^{r}}} \sum^{K}_{m=1}\varpi^m\sum_{i\in S_{m,r}}|i\rangle
\end{align}
where index $r$ takes the values in $\{1,\ldots,M-l_j\}$ and $l_j$ is the radial distance of $j$ from the center. The shells $S_{m,r}$, over which the linear combinations of position states are taken, are illustrated in Fig.~\ref{fig:cayley_states}(b). The number of these nonsymmetric states is: 
\be
\!(K-1)\!\sum^{M-1}_{l=1} (M-l)(K+1)K^{l-1}\!=\!(K+1)\!\left(\frac{K^M-1}{K-1}-M\right), 
\ee
where the factor $(K\,{+}\,1)K^{l-1}$ in the sum comes from the number of nodes in the $l$-th shell (possible seed nodes $j{\neq}0$), the multiplier $(K\,{-}\,1)$ corresponds to the possible choices of $\omega$ [initial states $\phi_0$ of the form~(\ref{eq:nonsymstate_betaneq0}) obeying Eq.~(\ref{eq:P_on_nonsym_states})], and $(M\,{-}\,l)$ is the number of nonsymmetric states in a given sector [states constructed by Eq.~(\ref{eq:propagation_forward})].

We can explicitly check that the constructed set of symmetric and nonsymmetric states is complete and
orthonormal.  Completeness, in particular, can be checked by calculating the total number of (non)symmetric states $\mathcal{N}$:
\begin{eqnarray}
\mathcal{N} 
&=& 1 + M + K M + (K + 1)\left(\frac{K^M-1}{K-1}-M\right)\nonumber \\
&=&1 + (K+1)\frac{K^M-1}{K-1} = 1 + (K+1)\sum_{l=0}^{M-1} K^l.
\end{eqnarray}
One can see that $\mathcal{N}$ matches the total number of nodes in a Cayley tree with connectivity $K$ and $M$ shells.
Therefore, any state $\ket{\Psi}$ is expanded using these basis states as
\begin{align}
\label{eqn:Psi-decomp}
    \ket{\Psi} =\sum \psi_l |l)+\sum_{j\in \mathcal{T}}\sum_{r}\sum_\varpi\psi_{j,r,\varpi}|r,\varpi)_j
\end{align}
where $\psi_0$ and $\psi_{j,r,\varpi}$ are the wave function components and $\mathcal{T}$ denotes the whole Cayley tree. 
In addition, we assume the dependence of the range of $\varpi$ on the seed node $j$ [cf.~Eq.~(\ref{eq:nonsymstate_betaneq0})] and the range of 
$r$ depends on the layer $l_j$ of the seed node. In the following text, we refer to eigenstates that can be expanded solely using (non)symmetric basis states as (non)symmetric eigenstates.

\subsubsection{Solution of BdG equations on Cayley trees}
\label{sec:BdG-on-Cayley}

In this subsection, we utilize 
the symmetry-adapted basis states to solve the
BdG equations on Cayley trees. 
We assume that the symmetry of the Caylee tree is not broken and that the order parameter $\Delta$ depends only on the distance from the central (root) point. These assumptions are
confirmed by exact diagonalization for smaller systems. Thereby, $\Delta_{i}\equiv\Delta_{l(i)}$, where $i$ is an index of a node, and $l(i)$ labels the shell containing node $i$. For simplicity, we will write $\Delta_l$ where appropriate.

A convenient way to use these symmetries is to apply the (non)symmetric basis states constructed in the previous section. In this symmetry-adapted basis, the Hamiltonian $h$ in Eq.~\eqref{eq:H_on_trees_posbasis} takes the block-diagonal form:
\begin{gather}
\label{eq:h_block_Cayley}
\tilde h = \begin{pmatrix}
 h_\textrm{sym} & 0 & 0 & 0 & \cdots \\
0 & h^{0}_\textrm{non-sym} & 0 & 0 & \cdots \\
0 & 0 & h^{1}_\textrm{non-sym} & 0 & \cdots \\
0 & 0 & 0 & h^{2}_\textrm{non-sym} & \cdots \\
\vdots & \vdots & \vdots & \vdots & \ddots \\
\end{pmatrix}
\end{gather}
where the individual non-symmetric blocks correspond to the various choices of the root node $j$ and of the initial wave function $|1,\omega)_j$, cf.~\cref{eq:nonsymstate_betaneq0}.
The transformation to (non)symmetric states also allows us to cast the BdG Hamiltonian in Eq.~\eqref{eq:BdG_Ham} to a block diagonal form. 

Recall that the BdG Hamiltonian acts on a direct sum of two single-particle sectors: the `electron' and the `hole' sector. 
We highlight this distinction by considering the vector $(u,v)^{T}$ of the components of the total wave function, as also formerly adopted in Eq.~\eqref{eq:BdG_Ham}. 
Applying rotation to the basis of (non)symmetric states, we introduce a vector of components in the new basis $(\tilde u, \tilde v)$ as follows:
\begin{gather}\label{eq:transfrom_nonsymstates}
  \begin{pmatrix}
    \tilde u\\ \tilde v
    \end{pmatrix}=\mathcal{U}\begin{pmatrix}
     u\\ v
    \end{pmatrix}=\begin{pmatrix}
    U&0\\0&U
    \end{pmatrix}    \begin{pmatrix}
    u\\v
    \end{pmatrix}
\end{gather}
where $U$ is the unitary matrix that rotates the components the in position basis into the components in the symmetry-adapted basis.
Let us point out that the chosen transformation $\mathcal{U}$ breaks the anticommutation properties of the particle/hole field operators; nevertheless, it preserves the form of the BdG Hamiltonian and enables the computation of the superconducting order parameter as we discuss below.

To proceed, recall that the order parameter $\Delta$ is constant on any shell, and that any (non)symmetric state is a linear combination of position states lying within 
some shell. 
It follows from these properties
that the diagonal matrix $\delta=\mbox{diag}(\Delta_1,\dots\Delta_\mathcal{N})$ commutes with $U$: $[\delta,U]=0$.
Therefore, the new BdG Hamiltonian $\tilde H_\textrm{BdG}=\mathcal{U} H_\textrm{BdG}\mathcal{U}^{\dagger}$ also takes the block-diagonal form:
\begin{gather}
\label{eq:Ham_block_Cayley}
\tilde H_\textrm{BdG} = \begin{pmatrix}
H_\textrm{sym} & 0 & 0 & 0 & \cdots \\
0 & H^{0}_\textrm{non-sym} & 0 & 0 & \cdots \\
0 & 0 & H^{1}_\textrm{non-sym} & 0 & \cdots \\
0 & 0 & 0 & H^{2}_\textrm{non-sym} & \cdots \\
\vdots & \vdots & \vdots & \vdots & \ddots \\
\end{pmatrix}. 
\end{gather}
The block structure means that in the basis of (non)symmetric states, it is sufficient to diagonalize each of the blocks separately. 
Each of the blocks takes the form of the BdG Hamiltonian:
 \begin{gather}
    \label{eqn:Block-BdG}
    H^\textrm{bl.}\begin{pmatrix}
    \tilde u^\textrm{bl.}_{n}\\\tilde v^\textrm{bl.}_{n}
    \end{pmatrix}=\begin{pmatrix}
    h^\textrm{bl.}-\mu&\mathbb{\Delta}^\textrm{bl.}\\\mathbb{\Delta}^\textrm{bl.}&-h^\textrm{bl.}+\mu
    \end{pmatrix}    \begin{pmatrix}
    \tilde u^\textrm{bl.}_{n}\\\tilde v^\textrm{bl.}_{n}
    \end{pmatrix}=E_{n}\begin{pmatrix}
    \tilde u^\textrm{bl.}_{n}\\\tilde v^\textrm{bl.}_{n}
    \end{pmatrix},
\end{gather}
where the superscript `$\textrm{bl.}$' denotes the choice of a (non)symmetric block of $\tilde H_{\textrm{BdG}}$, and the single-particle Hamiltonians $h^\textrm{bl.}$ correspond to the blocks of the Hamiltonian $\tilde h$ in Eq.~\eqref{eq:h_block_Cayley}. 
The diagonal matrix $\delta^\textrm{bl.}$ consists of the values $\Delta_l$ on those shells where the corresponding 
(non)symmetric sector has non-zero support. 

It should also be noted that while the transformation in Eq.~\eqref{eq:transfrom_nonsymstates} preserves values of $\Delta_l$, it does not preserve the structure of the BdG Hamiltonian in the second quantization picture, and therefore the self-consistent gap equation should be appropriately modified. 
The direct approach would be to plug $(\tilde u, \tilde v)^{T}$ given by the transformation in Eq.~\eqref{eq:transfrom_nonsymstates} into the equation~\eqref{eq:selfcons_delta}. The gap equation written in terms of the components of the wave function in the symmetry-adapted basis then reads:
\be
\Delta_i=\frac{U}{2}\sum_{n} (U^{\dagger}\tilde u_n)_i (U^{\dagger}\tilde v_n)^{*}_i \tanh(\frac{E_n}{2 T}),
\ee
where the sum is going over all eigenstates in $H_{\textrm{BdG}}$. 
However, working with the obtained expression in the general form is rather demanding. 
Therefore, using the radial symmetry of the Cayley trees, we apply the transformation in several steps, eventually arriving at the final form of the gap equation listed in Eq.~\eqref{eq:selfcons_Cayley}.

The crucial feature of symmetric and nonsymmetric states is the location of their support on the Cayley tree. 
While states of the symmetric sector
can take non-zero values everywhere on the tree, states constituting a chosen nonsymmetric sector have non-zero values only on the branches emanating from the corresponding seed node of the tree. 
Since the Cayley tree is radially symmetric, it is not necessary to consider all nonsymmetric blocks; rather, 
we can choose a subset
of the nonsymmetric BdG Hamiltonian blocks
$H^{(l_j,\varpi)}_\textrm{non-sym}$ such that their supports have non-zero intersection. In contrast to Eq.~(\ref{eq:Ham_block_Cayley}), we are here specifying the Hamiltonian blocks
by the choice of $l_j$ (the distance of the `seed' node $j$ from the center) and the root of unity $\varpi$. 
For the following considerations, it turns out to be sufficient to choose blocks whose seeds lie on a `ray' connecting an arbitrary boundary site to the central node.

The subsequent algorithm for calculation of the order parameter profile is thus summarized as follows. 
First, we select nodes lying on a ray from the central site to one of the boundary sites. 
For each node $j$ in this selection, we consider the corresponding (non)symmetric basis states generated from the node and diagonalize only the corresponding blocks of the BdG Hamiltonian: $H_\textrm{sym}$ and $H^{(l_j,\varpi)}_\textrm{non-sym}$, where $l_j$ takes values in the range $\{0,1,2,\dots, M-1\}$, and possible values of $\varpi$ depend on $l_j$ as in \cref{eq:nonsymstate_betaneq0,eq: nonsymm basis l=0,eq: nonsymm basis}.
The mixing between solutions of the blocks is given by self-consistent equations, which require the knowledge of the eigenstates of the (non)symmetric BdG blocks:
\begin{gather}
    \Delta_i=\frac{U}{2}\sum_{n}u^\textrm{sym.}_{n,i} (v^\textrm{sym.})^{*}_{n,i} \tanh(\frac{E_n}{2 T})\nonumber \\
    +\frac{U}{2}
    \sum_{ \substack{\alpha \in \textrm{ nonsymmetric blocks}\\\textrm{with seed}\,\, j\in \mathcal{P}(i)}}
    \sum_{n}u^{\alpha}_{n,i} (v^{\alpha})^{*}_{n,i} \tanh(\frac{E_n}{2 T}). 
    \label{eq:gapeq_posbasis}
\end{gather}
By $\Delta_i$ we denote the order parameter at site $i$, and with the range $\mathcal{P}(i)$ we mean the nodes that are parents of site $i$. 
The summation over $n$ indicates the summation over eigenstates of the corresponding (non)symmetric blocks of the BdG Hamiltonian; therefore, the specific range of the index $n$ depends on the choice of a block (we omit this dependence in the gap equation). The choice of a (non)symmetric block is indicated by the superscripts $\textrm{sym.}$ or $\alpha$.
Note that in Eq.~(\ref{eq:gapeq_posbasis}) we still write the eigenstates of BdG (nonsymmetric) blocks in the original position basis. Therefore, as the next step, we express the self-consistent equations in the basis of symmetric and nonsymmetric states.

To achieve the desired change of basis, we need to extract the single-particle parts of the considered BdG blocks.
Applying the tight-binding Hamiltonian $h$ to symmetric states, one can obtain the Hamiltonian in the symmetric sector~\cite{Mahan-2001-PRB}:
\begin{gather}\label{eq:H_sym}
    h_\textrm{sym} = \begin{pmatrix}
0 & \sqrt{K+1} & 0 & 0 & \cdots \\
\sqrt{K+1} & 0 & \sqrt{K} & 0 & \cdots \\
0 & \sqrt{K} & 0 & \sqrt{K} & \cdots \\
0 & 0 & \sqrt{K} & 0 & \cdots\\
\vdots & \vdots & \vdots & \vdots & \ddots \\
\end{pmatrix}.
\end{gather}
In a similar way, one can obtain the Hamiltonian for nonsymmetric sectors \cite{Aryal-IOP-2020, Hamanaka:2024}:
\begin{gather}\label{eq:H_nonsym}
    h^{l_j}_\textrm{non-sym} = \begin{pmatrix}
0 & \sqrt{K} & 0 & 0 & \cdots \\
\sqrt{K} & 0 & \sqrt{K} & 0 & \cdots \\
0 & \sqrt{K} & 0 & \sqrt{K} & \cdots \\
0 & 0 & \sqrt{K} & 0 & \cdots \\
\vdots & \vdots & \vdots & \vdots & \ddots \\
\end{pmatrix},
\end{gather}
where $l_j$ is the distance of the seed of the nonsymmetric states from the central node, and the size of the matrix is $(M\,{-}\,l_j{-}1)\times (M\,{-}\,l_j{-}1)$.
In total, we need to diagonalize $(M\,{+}\,1)$ matrices, where $M$ is the number of shells in the Cayley tree surrounding the central node.
Of these $(M\,{+}\,1)$ matrices, one comes from the symmetric block, and $M$ come from nonsymmetric states (recall that the seed of nonsymmetric states can lie at the root node, but not in the outermost layer).

\begin{figure*}
\includegraphics[width=\linewidth]{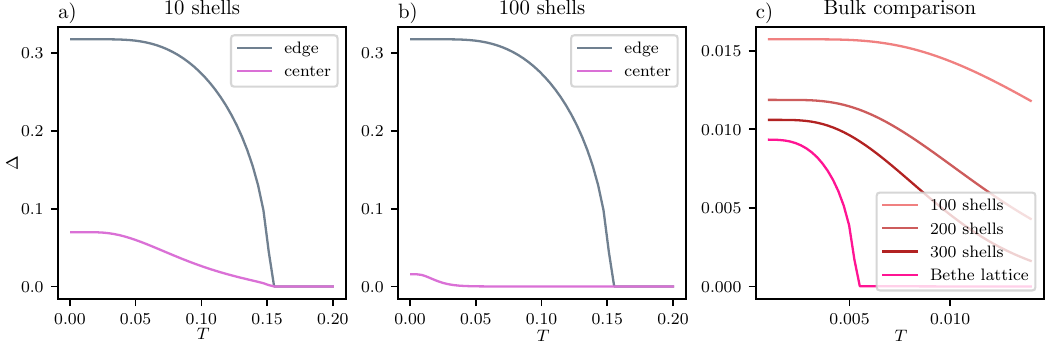}
\caption{Superconducting order parameter of
finite and infinite Cayley trees with vertex degree $K+1=3$ for parameters $\mu=0$ and $U=1$. We display the computed order parameter at the central (root) site and at the edge sites for (a) small system with $10$ shells, and for (b) larger system with $100$ shells.
The right panel compares the value of the order parameter at the center of trees with various radial sizes against
the thermodynamic limit of the Bethe lattice. }
    \label{fig:finite_tree_phases}
\end{figure*}

To correctly utilize the basis of (non)symmetric states, we further need to consider two effects: the degeneracy of the nonsymmetric states with the same `seed' node, and the exponential decay of the (non)symmetric states written in the position basis. 
First, the degeneracy of nonsymmetric states is equal to the number of successor nodes of the `seed' node minus one:
\be
\textrm{deg}(l_j)=\left\{\begin{array}{ll}
K& \textrm{for $l_j = 0$,}\\K-1 & \textrm{for $l_j >0$.}
\end{array}\right.
\ee
After taking into account these multipliers, we obtain: 
\begin{gather}
    \Delta_l=\frac{U}{2}\bigg[\sum_{n} u^\textrm{sym.}_{n,l} (v^\textrm{sym.}_{n,l})^* \tanh(\frac{E_n}{2 T}) \nonumber \\
    \label{eq:selfcons_Cayley_inter}
    {}+K\sum_{\substack{\textrm{nonsymmetric}\\\textrm{block with } l_j=0}} \sum_{n}  u^{0}_{n,l} (v^{0}_{n,l})^* \tanh(\frac{E_n}{2 T}) \\
    \nonumber
    {}+(K-1)\sum_{\substack{\textrm{nonsymmetric}\\\textrm{blocks with } l_j>0}} \sum_{n}  u^{l_j}_{n,l} (v^{l_j}_{n,l})^* \tanh(\frac{E_n}{2 T})\bigg].
\end{gather}
Here, in contrast with Eq.~\eqref{eq:gapeq_posbasis}, we work only with shell indices, simplifying the equation a bit further, with $l$ denoting the index of the shell.

Second, to finalize the transformation to the components $(\tilde u,\tilde v)^T$ in the symmetry-adapted basis, we need to consider the multipliers coming from the normalization coefficients of (non)symmetric states as in \cref{eq: symm basis,eq:nonsymstate_betaneq0,eq: nonsymm basis l=0,eq: nonsymm basis}. That gives us additional multipliers for $(u,v)^T$ in different (non)symmetric blocks in Eq.~\eqref{eq:selfcons_Cayley_inter}. Since the matrix $U^{\dagger}$ expressing  $(u,v)$ via $(\tilde u, \tilde v)^T$ is also the matrix relating the (non)symmetric basis states with position basis states, the additional multipliers for $(u,v)^T$ are the normalization coefficients exactly.
Taking this effect into account, we obtain the final form of the self-consistent equations for the order parameter written via the components in the basis of (non)symmetric states $(\tilde u, \tilde v)^T$:
\begin{gather}
    \Delta_l=\frac{U}{2}\bigg[\sum_{n; l\ge 0}N^{0}_l \tilde u^{\textrm{sym}}_{n,l} \tilde v^{\textrm{sym}}_{n,l} \tanh(\frac{E^0_n}{2 T}) \nonumber \\
    \label{eq:selfcons_Cayley}
    {}+\frac{K^2}{K+1}\sum_{n;l>0} K^{-l}\tilde u^{0}_{n,l} \tilde v^{0}_{n,l} \tanh(\frac{E^1_n}{2 T}) \\
    \nonumber
    {}+(K-1)\sum_{n; l> l_j> 0}K^{l_j-l}\tilde u^{l_j}_{n,l} \tilde v^{l_j}_{n,l} \tanh(\frac{E^k_n}{2 T})\bigg],
\end{gather}
where we distinguish different nonsymmetric blocks by $l_j$, which is the distance of the `seed' from the center.
The sum over $n$ denotes the summation over eigenstates of (non)symmetric blocks, so it implicitly depends on $l_j$.
Eigenenergies $E^k_n$ and eigenfunctions $(\tilde u^{\textrm{sym}}_{n},\tilde u^{\textrm{sym}}_{n})^{T}$ and $(\tilde u^{l_j}_{n},\tilde u^{l_j}_{n})^{T}$ are calculated separately for each (non)symmetric block of the BdG Hamiltonian $H_{\textrm{BdG}}$, and since the blocks are real-valued, we choose the real gauge for $(\tilde u, \tilde v)^T$ and omit complex conjugation.
The weight $N^0_i$ used for the symmetric states is defined as:
\begin{gather}
    N^0_l=\begin{cases}
        1, \quad\textrm{if}\quad l=0\\
        \frac{1}{(K+1)K^{l-1}} , \quad\textrm{if}\quad l>0.
    \end{cases}
\end{gather}

\subsubsection{Results}
\label{sec:Cayley-results}

 \begin{figure}
    \centering
    \includegraphics[width=\linewidth]{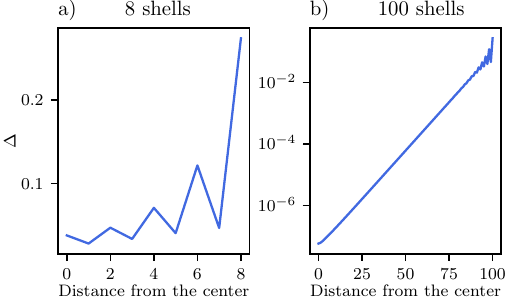} 
    \caption{The spatial profile of the order parameter for the Caylee tree with connectivity $q=2$, with the total number of layers $M=8$ (left panel) and $M=100$ (right panel). The parameters are $U=1$, $\mu=0$ and $T=0.1$. The profile of the order parameter on the right panel is shown on a logarithmic scale.}
    \label{fig:Delta_Caylee_tree}
\end{figure}

The presented calculation scheme allows us to calculate the self-consistent order parameter on trees with radial size up to a few hundred shells. 
The importance of the large radial size 
is demonstrated by Fig.~\ref{fig:finite_tree_phases}, where the slice of the phase diagram for trees with different numbers of layers is shown. 
The figures compare the order parameter at the central node and the order parameter at the boundary shell for $U=1$ and $\mu=0$. 
One can see that for a relatively small radial size [$10$ shells, Fig.~\ref{fig:finite_tree_phases}(a)], the boundary superconducting gap is enhanced. 
However, the critical temperature for the bulk and the boundary is the same. 
To observe the emergence of the second critical temperature, one has to consider larger trees [$100$ shells, Fig.~\ref{fig:finite_tree_phases}(b)]. 
It is interesting to note that the phase diagram for the boundary superconductivity does not change noticeably from smaller to larger trees. 
The superconductivity at the center of a tree, in turn, slowly approaches the thermodynamic limit with the increase of the radial size, as demonstrated by Fig.~\ref{fig:finite_tree_phases}(c). 
Nevertheless, even for the tree with $100$ shells (total number of sites is of the order $10^{24}$), the bulk critical temperature remains noticeably higher than the critical temperature of the thermodynamic limit.
A typical profile of the order parameter localized at the boundary, plotted in a log-scale, is shown in Fig.~\ref{fig:Delta_Caylee_tree}(b). The profile is calculated for a Cayley tree with vertex degree $K+1=3$ and with $M=100$ 
layers for parameters $U=1$, $\mu=0$, and $T=0.1$. 
One can clearly observe the dichotomy
between the bulk, where the superconducting gap is absent, and the boundary, which hosts an exponentially localized superconducting gap.

\begin{figure}
    \centering
    \includegraphics[width=\linewidth]{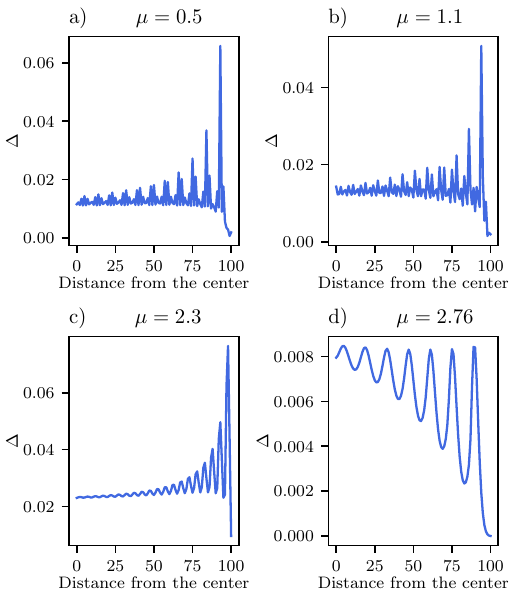} 
    \caption{The spatial profile of the order parameter for the Caylee tree with connectivity $K=2$, with the total number of layers $M=100$ for several values of chemical potential $\mu$. The parameters are $U=1$ and $T=0.001$.}
    \label{fig:Delta_Caylee_tree_varmu}
\end{figure}

One can also study how the picture changes when varying the chemical potential $\mu$. 
In the considered case of $\mu=0$, we observed that the order parameter is higher exactly at the edge. 
However, for other chemical potentials, the picture can be different, which can be justified as follows. For BCS-like theories, the critical temperature is an increasing function of the density of states (DOS). 
The fact that $T_\textrm{c}$ at the boundary is higher than in the bulk for $\mu = 0$ should be related to the fact that the local density of states (LDOS) at $\mu=0$ is larger near the edge than the bulk DOS. 
However, the integral of LDOS over energy is
the same at each site, meaning that there should be some other value of $\mu$ where LDOS at the boundary is \emph{lower} than the bulk DOS. 
Therefore, for such a choice of $\mu$,
we should expect the system to become superconducting only once we cool the system to the bulk $T_\textrm{c}$, with the order parameter possibly suppressed near the boundary.

To verify this prediction, we show in Fig.~\ref{fig:Delta_Caylee_tree_varmu} the profile of the superconducting order parameter for several values of the chemical potential $\mu$ for very low temperature $T=0.001$.
One can see that for all chosen values $\mu \neq 0$, the order parameter exactly at the boundary is close to $0$.
Nevertheless, an increase in $\Delta$ towards the boundary is present for three of the plots [Fig.~\ref{fig:Delta_Caylee_tree_varmu}(a,b,c)], although with oscillations whose wavelength is larger than the period-2 oscillations observed for $\mu=0$. In contrast, in Fig.~\ref{fig:Delta_Caylee_tree_varmu}(d), where we set the chemical potential close to the spectral edge ($E_\textrm{edge}\approx 2.828$). The order parameter does not increase towards the boundary; rather, averaging over the oscillations results in dampening of $\Delta$ towards the boundary.
These results show that the phenomenon observed for $\mu{=}0$, where there is a distinguished boundary-superconductivity phase, is subject to an appropriate choice of the chemical potential~$\mu$.

\begin{figure}
\includegraphics[width=\linewidth]{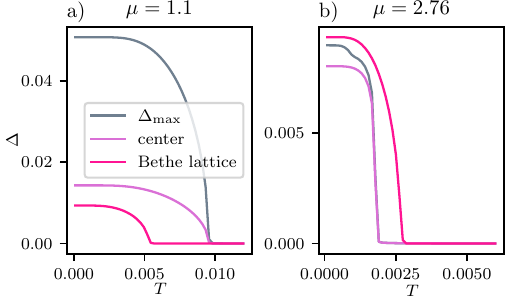}
\caption{The slice of the phase diagram for finite Cayley trees with $K=2$ calculated for the order parameter at the center and its maximum value for two chemical potentials $\mu=1.1$ and $\mu=2.76$.  The interaction potential is $U=1$, and the number of shells equals $100$. The order parameter in the thermodynamic limit of Bethe lattice is shown for comparison.}
    \label{fig:finite_tree_phases_nonzeromu}
\end{figure}

To further understand the difference between the boundary and the bulk superconductivity, we compare the maximum value of $\Delta$ and its value at the center of the Cayley tree with $K+1=3$, plotted as a function of temperature, for two different chemical potentials: $\mu = 1.1$ [Fig.~\ref{fig:finite_tree_phases_nonzeromu}(a)] and $\mu=2.76$ [Fig.~\ref{fig:finite_tree_phases_nonzeromu}(b)].  
The Hubbard potential is $U=1$ and we assume $M=100$ layers of the tree.
For both choices of $\mu$, we also compare the results with the thermodynamic limit of the Bethe lattice. 
It is visible in Fig.~\ref{fig:finite_tree_phases_nonzeromu}(a)
that the maximum value of $\Delta$ and the value of $\Delta$ at the center exhibit
the same critical temperature, despite the value of 
$\Delta$ at the center being noticeably lower. 
Since the thermodynamic critical temperature ($T_\textrm{c} \approx 0.005$) is less than the critical temperature observed at the center of a system with $M=100$ sites ($T_\textrm{c} \approx 0.01$), we infer that even such a large radial size is insufficient to exhibit the bulk behavior.
Nevertheless, we anticipate that a convergence of the superconducting gap at the center of the Cayley tree to the Bethe lattice prediction would eventually be observed for values of $M$ further exceeding those considered in our work. 
We hypothesize that the absence of convergence of the critical point at the center to the thermodynamic value of $T_\textrm{c}$ for the large assumed system size ($M=100$) may be related
to the larger oscillation length at finite $\mu \neq 0$ compared to the period-2 oscillations present for $\mu=0$; however, a deeper analysis would be necessary to definitely settle this question.

In contrast, for chemical potential set to $\mu = 2.76$ [Fig.~\ref{fig:finite_tree_phases_nonzeromu}(b)] we observe that the maximum value of $\Delta$ is \emph{less} than the bulk value in the thermodynamic limit. Therefore, one can assume that for this choice of $\mu$, the value of $\Delta$ at the center converges to the thermodynamic-limit value of $\Delta$ from below upon increasing the system size. 
Such interpretation appears compatible with the observation of boundary-suppressed superconductivity in Fig.~\ref{fig:Delta_Caylee_tree_varmu}(d).

The dependence of the rate of convergence can be explained, if one carefully investigates the local density of states (LDOS) for a given energy $\epsilon$.
Thus, one needs to find which states contribute to LDOS at fixed $\epsilon$. The contributions can come from states of various (non)symmetric blocks. 
However, for large trees, the contribution of symmetric states is negligibly small, and we can focus solely on nonsymmetric states. The energies of a nonsymmetric block of length $L$ are:
\be
\label{eqn:Cayley-energetics}
\epsilon_{p,L}=2\sqrt{K}\cos(\pi\kappa_{p,L}),
\ee
where $p\in\{1,2,\ldots, L\}$, and $\kappa_{p,L}$ is a (quasi)momentum defined as $\kappa_{p,L}=\frac{p}{L+1}$.

For each site at the shell $l$, there exist $M-l$ nonsymmetric blocks with lengths larger than $L$ that could contribute to the LDOS at the site. 
At fixed energy $\epsilon$, the contributions from nonsymmetric states of different blocks are added when the states have exactly the same energy $\epsilon$. 
This occurs when the corresponding momenta $\kappa$ in different nonsymmetric blocks of lengths $L$ and $L'$ are the same:
\be
\kappa=\frac{p}{L+1}=\frac{p'}{L'+1}.
\ee
Since we consider finite trees (potentially including immensely large ones), the admissible momenta are rational numbers lying in the interval $(0,1)$, since $L$ and $p\le L$ are positive integers. 
Each rational number can be represented as a ratio between two coprime numbers. 
Hence, for each admissible momentum $\kappa$, we can find minimum values of $p$ and $L$, such that $p_\textrm{min}$ and $L_\textrm{min}+1$ are coprime. By multiplying $p_\textrm{min}$ and $L_\textrm{min}+1$ by the same integer number, we obtain all the other possible values of $L$ and $p$. 
In particular, lengths of the nonsymmetric blocks that exhibit states with momentum $\kappa = p_\textrm{min}/(L_\textrm{min}+1)$ can be expressed as $L=d(L_\textrm{min}+1)-1$, where $d$ is a positive integer.

It follows from the above considerations that the shorter the minimal length $L_\textrm{min}$, the more states contribute to the local density of states, and the value of the order parameter should be higher. 
Moreover, one can notice that the seed nodes of nonsymmetric blocks contributing to LDOS at the given energy are equidistant from each other. That allows us to introduce the characteristic length of oscillations $\ell=L_\textrm{min}+1$. 
Summing up the discussion, we anticipate that the shorter the characteristic length of oscillations $\ell$, the higher the maximum value of $\Delta$, and the faster the convergence to the explicit distinction between the bulk and the boundary phases.

For instance, the fact that all nonsymmetric blocks with odd length ($L_\textrm{min}=1$) have the eigenvalue $E=0$ suggests that the boundary critical temperature should be the highest for $\mu=0$. 
This observation also agrees with the period-$2$ oscillations of the profile of the order parameter shown in Fig.~\ref{fig:Delta_Caylee_tree}.
The next maximum of boundary density of states, 
characterized by oscillations with period $\ell=3$, occurs at energies $E=\pm\sqrt{K}$. 
The energies corresponding to characteristic length $\ell=4$ are $E=\pm\sqrt{2K}$. 
In general, we find from Eq.~(\ref{eqn:Cayley-energetics}) that further (and consecutively lower) maxima of the boundary density of states, associated with oscillations with integer period $\ell$, occur at energies $E=2\sqrt{K}\cos{(p\pi/\ell)}$ where $p$ is coprime with $\ell$.

\begin{figure}
\includegraphics[width=\linewidth]{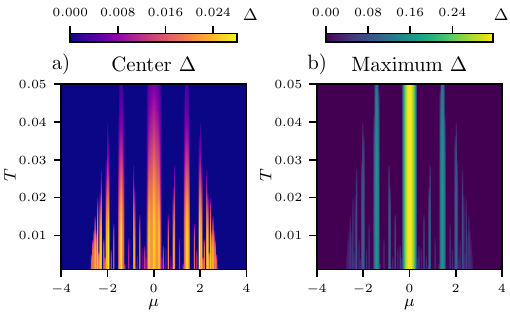}
\caption{Superconducting order parameter $\Delta$
calculated for $U=1$ at the center (a) and its maximum value (b) for the Cayley tree with vertex degree $K+1=3$ and with $M=40$ layers. 
For a better representation, the results are shown in different color schemes. 
The panels demonstrate that the $\Delta$ at the center and $\Delta_\textrm{max}$ follow a similar pattern; however, the values at the center, in general, are significantly lower.
Observe also that the onset of superconductivity in the center of the system with $40$ layers occurs at a significantly larger temperature than in the bulk (Bethe lattice), cf.~Fig~\ref{fig:bethe_example}.
}
\label{fig:cayley_tree_phase_diags}
\end{figure}

To verify the prediction on enhanced boundary superconductivity at the specified energies, we show in Fig.~\ref{fig:cayley_tree_phase_diags} the phase diagram in variables $(\mu,T)$ calculated for $M=40$. The figure demonstrates that the peaks of maximum $\Delta$ are located at the energies given by Eq.~\eqref{eqn:Cayley-energetics} with rational $\kappa$ which is a ratio of small coprime numbers. 
Higher values of $\Delta$ correspond to energies with lower integers $\ell$.
Accordingly, the cases with low $\ell$ also demonstrate noticeably enhanced disparity between the maximum value of $\Delta$ vs.~the value of $\Delta$ at the center of~the~tree. 

With an increase in the radial size, the order parameter at the center converges to the thermodynamic limit of the Bethe lattice shown in Fig.~\ref{fig:bethe_example}. At the same time, one can expect that the maximum values of $\Delta$ are stable with the increase of radial size, so that the presence of boundary superconductivity for large trees can be determined from the trees of lower radial size. 
Notably, if some value of $\mu$ and $T$ displays a non-zero maximum $\Delta$ in the Cayley tree calculation [Fig.~\ref{fig:cayley_tree_phase_diags}(b)] but a zero bulk $\Delta$ in the Bethe-lattice calculation (Fig.~\ref{fig:bethe_example}), then these parameters correspond to the phase with boundary-only superconductivity. 
Nevertheless, the convergence to the thermodynamic limit appears to depend sensitively on the choice of the chemical potential $\mu$. 
In particular, one can observe the pronounced peak for $\Delta$ at the central site for $\mu=0$; in contrast, the value of $\Delta$ in the thermodynamic limit exhibits a local \emph{minimum} at this choice of $\mu$.

\subsubsection{Relation to density of states on Cayley trees}
\label{sec:Cayley_ldos}

To complete the discussion of superconducting order in Cayley trees, we supplement
the presented arguments with analytical calculations of LDOS on Cayley trees. 
To simplify the analysis, we count the shells starting from the boundary as $r=M-l+1$, so that index $r=1$ corresponds to the last shell of a tree. 
In the limit of large trees, $M\gg1$, the contribution of symmetric states is negligible, so we are interested only in the eigenstates coming from the nonsymmetric blocks. 
In the basis of nonsymmetric states, they are related to 
eigenstates of a 1D chain with open boundaries:
\begin{gather}
\tilde\psi(r)_{L, p}=\begin{cases} \sqrt{\frac{2}{L+1}}\sin\frac{\pi p r}{L+1}, \quad 1\le r\le L,  \\0, \quad r> L,\end{cases}
\end{gather}
where index $r$ measures the distance of a site from the boundary, $n=1,2,\ldots,L$, and $L$ is the number of states in the nonsymmetric block, which is also equal to the distance of the `seed' from the boundary. 

After rotating to the position basis and neglecting the phase coming from the nontrivial root of unity, the states acquire an exponential prefactor:
\begin{gather}\label{eq:Cayley_tree_eigstate}
\psi(r)_{L, p}=\begin{cases} \sqrt{\frac{2K^{-L}}{L+1}}K^{\frac{r}{2}}\sin\frac{\pi p r}{L+1}, \quad 1\le r\le L,  \\0, \quad r> L.\end{cases}
\end{gather}
The LDOS at the given site is written as:
\be
\textrm{LDOS}(\epsilon,r)=\sum_n |\psi_n|^2\delta(\epsilon-\epsilon_n),
\ee
where the sum is taken over all energies. Taking a particular energy $\epsilon_{\ell,p_\textrm{min}}$ corresponding to rational momentum $\kappa={p_\textrm{min}}/{\ell}$, we can write its weight in the local density of states:
\begin{eqnarray}
& \phantom{=}& w_{\ell,p_\textrm{min}}(r) = (K-1)\sum^{d_\textrm{max}}_{d=1} |\psi(r)_{d \ell-1, d p_\textrm{min}}|^2=\nonumber\\
& = & \sqrt{\tfrac{2}{\ell}} \sum^{d_\textrm{max}}_{d=1} \theta(d\ell-1-r)\tfrac{K^{r-d\ell+1}}{d}\sin^2\tfrac{\pi p_\textrm{min} r}{\ell},
\label{eq:ldos_delta_weight}
\end{eqnarray}
where $\theta(x)$ is Heaviside step function, and the maximal value of $d$ is determined by the size of the Cayley tree. The multiplier $K-1$ comes from the nonsymmetric blocks corresponding to different roots of $1$.
The LDOS is given by the sum over all admissible energies scaled with the corresponding weights:
\be
    \! \textrm{LDOS}(\epsilon,r)\! = \!\! \sum^{M}_{\ell=2} \!\!\! \sum^{\textrm{max}_{\ell}(p_\textrm{min})}_{\substack{p_\textrm{min}=1\\ \textrm{gcd}(p_\textrm{min},\ell)=1}} \!\!\!\!\!\!\!\!\! w_{\ell,p_\textrm{min}}\!(r) \delta \! \left(\epsilon \! - \! 2\sqrt{K}\cos{\!\tfrac{\pi p_\textrm{min}}{\ell}}\!\right)\!,\!
\ee
where with $\textrm{max}_{\ell}(p_\textrm{min})$ we denote the maximum value of $p_\textrm{min}$ for a given $\ell$.

To proceed, let us fix a distance from the boundary $r$, and we estimate $w_{\ell,p_\textrm{min}}(r)$ when the size of a tree tends to infinity ($M\to\infty$). 
Then we can write:
\be
\lim_{M\to\infty} w_{\ell,p_\textrm{min}}(r)=(K-1)\sqrt{\tfrac{2}{\ell}}K^{r+1}\sin^2\tfrac{\pi p_\textrm{min}r}{\ell} \!\!\!\! \sum^{\infty}_{d=\lfloor\frac{r+1}{\ell}\rfloor+1} \!\!\!\! \tfrac{K^{-d\ell}}{d},
\ee
where $\lfloor x \rfloor$ is the floor function. 
One can see that whenever $r$ is not a multiple of $\ell$, the limit $\lim_{M\to\infty} w_{\ell,p_\textrm{min}}(r)$ is always finite. 
Therefore, LDOS on sites closer to the boundary does not converge to a smooth function, but accumulates more and more $\delta$-functions with small but non-vanishing weights. The limit LDOS becomes an infinite sum of $\delta$-functions with non-trivial support on energies that correspond to
rational momenta.

Thereby, we have found that LDOS at the boundary has a completely different form than DOS in the bulk, which is a smooth function. This result gives an additional argument for the existence of a separate boundary phase on Cayley trees and explains the stability of the maximum values of $\Delta$ in the preceding calculations. 
In turn, the existence of $\delta$-peaks surrounded by the $\delta$-peaks with smaller weight in the boundary LDOS is the reason why, for some chemical potentials, the difference of the bulk and boundary critical temperatures can amount to more than an order of magnitude.

\section{BCS theory on continuous hyperbolic spaces}\label{sec:cont_spaces}

In this section, we leave behind the case of discrete lattices and focus instead on developing the BCS mean-field formulation in \emph{continuous} hyperbolic spaces.
Our discussion is structured as follows.
We first treat in Sec.~\ref{sec:uni_space} the uniform hyperbolic plane in the absence of boundaries, exploiting its isometries to obtain the single-particle Green’s function and density of states (DOS). 
From these, the gap equation follows in a form identical to flat space except that curvature enters through the modified DOS function. 
We then introduce in Sec.~\ref {sec:hyperspace} semi-infinite systems by considering regions bounded by a horocycle and a geodesic, and we study their LDOS and eigenstates.
Finally, we numerically solve the self-consistent equations, showing the existence of thin-film superconductivity with a higher critical temperature than in the bulk.

\subsection{Uniform space}\label{sec:uni_space}

We start from the \textit{ab intio} continuous description of the problem; we do it in the whole hyperbolic plane in order to study bulk superconducting properties. Clearly, translational invariance and isotropy require normal and anomalous thermal Green's functions to be dependent on the hyperbolic distance between two points only, i.e. 
\begin{eqnarray}
    \mathfrak{G}(\tau;x,x^\prime)&=&\mathfrak{G}(\tau;d(x,x^\prime)) \\ \mathfrak{F} (\tau;x,x^\prime)&=&\mathfrak{F}(\tau;d(x,x^\prime)),
    \end{eqnarray}
    where $\mathfrak{G}$ and $\mathfrak{F}$ are the normal and the anomalous thermal Green's functions, $\tau$ is imaginary time, $x,x'$ are coordinates in the two-dimensional hyperbolic plane, and $d(x,x')$ denotes the distance between points with coordinates $x$ and $x'$.

If the pairing occurs in the s-channel, the equation of motion for the Green's functions
reads \cite{Abrikosov:107441}:
\begin{widetext}
	\begin{gather} \label{Gorkov_eq}
		\begin{pmatrix}
			\left\{-\frac{\partial}{\partial \tau} + \frac{\triangle_{\mathbb{H}^2}}{2m}+\mu\right\} & \Delta\\
			\Delta^* & \left\{\frac{\partial}{\partial \tau} + \frac{\triangle_{\mathbb{H}^2}}{2m}+\mu\right\}
		\end{pmatrix}
		\Bigg(
		\begin{matrix}
			\mathfrak{G}(\tau;x,x^\prime) &  \mathfrak{F}(\tau;x,x^\prime) \\
			\mathfrak{F}^*(\tau;x,x^\prime) & - \mathfrak{G}(\tau;x^\prime, x)
		\end{matrix}
	\Bigg) = \Bigg(
 		\begin{matrix}
			\delta(\tau)\delta(x,x^\prime) &  0 \\
			0 & \delta(\tau)\delta(x,x^\prime)
		\end{matrix}
  \Bigg) 
	\end{gather}
    \end{widetext}
	where $\Delta=U\lim_{\tau\to 0+}\mathfrak{F}(\tau;0,0)$. In the previous equation, $\triangle_{\mathbb{H}^2}$ is the Laplace-Beltrami operator on hyperboloid, and $\delta(x,x^\prime)$ represents the hyperbolic delta-function which should be understood in the following way: 
    \begin{equation}
    \int_{\mathbb{H}^2} \textrm{d}^2 x \sqrt{g} \cdot \delta(x,x^\prime) f(x) = f(x^\prime),    
    \end{equation}
    where $g$ is the determinant of the metric tensor. 
    For instance, in hyperbolic polar coordinates $(\varrho,\phi)$,
    the delta function localized at zero can be represented in terms of the `conventional' Dirac delta function as  $\delta(\varrho)/2\pi\sinh(\varrho)$.
    
    In hyperbolic polar coordinates, the Laplace-Beltrami operator takes the form:
	\begin{equation}
    \label{eqn:H2-Laplace-Beltrami}
		\triangle_{\mathbb{H}^2} = \frac{1}{\sinh \varrho} \frac{\partial}{\partial \varrho} \sinh \varrho \frac{\partial}{\partial \varrho} + \frac{1}{\sinh^2 \varrho} \frac{\partial^2}{\partial \phi^2},
	\end{equation}
    where, for convenience, we set the curvature radius of the hyperbolic space as one: $R=1$.
	The eigenfunctions of the Laplace-Beltrami operator are 
    \be
    \label{eqn:LB-eigenstates}
    \psi_{m,k}(\varrho, \phi)= e^{i m \phi} P^m_{-\frac{1}{2}+ik}(\cosh(\varrho))
    \ee
    where $P^m_{-\frac{1}{2}+ik}(\cosh(\varrho))$ is the conical (Mehler) function. In the case of $m{=}0$, we will omit the superscript and write $P_{-\frac{1}{2}+ik}$. The eigenvalues corresponding to the sector with $m{=}0$ are ${-}{k}^2{-}1/4$. For future considerations, we also introduce quantity $\epsilon_k$, defined as $\epsilon_k\equiv(k^2+\frac{1}{4})/2m$.
    In Eq.~(\ref{Gorkov_eq}), one can set $x^\prime$ to be zero, as the solution for non-zero $x^\prime$ can be easily obtained from homogeneity and isotropy. Moreover, from isotropy, Green's functions depend only on the radial coordinate $\varrho$:
	\begin{equation}
		\mathfrak{G}(\tau;x,0)=\mathfrak{G}(\tau,\varrho) \quad \textrm{and}\quad \mathfrak{F}(\tau;x,0)=\mathfrak{F}(\tau,\varrho).
	\end{equation} 
	Expanding Green's functions into their Matsubara and Mehler (see appendix \ref{app:MF_sec}) components, one finds:
	\begin{align}
		\mathfrak{G}(\tau,\varrho)&=T\sum_n e^{-i\omega_n\tau}\int^\infty_0 dk \cdot \psi_k(\varrho) \mathfrak{G}(\omega_n,k),\\
		  \mathfrak{F}(\tau,\varrho)&=T\sum_n e^{-i\omega_n\tau}\int^\infty_0 dk \cdot \psi_k(\varrho) \mathfrak{F}(\omega_n,k)
	\end{align}
    where $T$ is temperature, the Matsubara frequencies have the values $\omega_n=(2n+1) \pi T$, and $\mathfrak{G}(\omega_n,k)$ is the Mehler-Fock transform of $\mathfrak{G}(\omega_n,\varrho)$. 
 
    Defining $\xi_k \equiv \epsilon_k-\mu$ as the one-particle eigenenergies counted from the Fermi level (and assuming that both the gap $\Delta$ and the anomalous function $\mathfrak{F}$ are real-valued), Eq.~\eqref{Gorkov_eq} reduces to:
    \begin{equation} 
        \label{eqn:cont-greeens}
        \begin{array}{rcc}
        (i\omega - \xi_k) \mathfrak{G}(\omega,k) + \Delta \mathfrak{F}(\omega,k) & = & \frac{k\tanh(\pi k)}{2\pi} 
        \\
        (i\omega + \xi_k) \mathfrak{F}(\omega,k) + \Delta \mathfrak{G}(\omega,k) & = & 0
        \end{array} 
    \end{equation}
    which provides a solution:
	\begin{align}
		\label{eqn:cont-thermal-G-Green}
        \mathfrak{G}(\omega,k)&=-\frac{k\tanh(\pi k)}{2\pi}\frac{i\omega+\xi_k}{\omega^2+\xi_k^2+\Delta^2}\\
		\mathfrak{F}(\omega,k)&=\frac{k\tanh(\pi k)}{2\pi}\frac{\Delta}{\omega^2+\xi_k^2+\Delta^2}
	\end{align}
	The value of the superconducting gap can now be found from the self-consistency condition $\Delta=U\mathfrak{F}(0+,0)$, which gives:
	\begin{equation}
		1=UT\sum_n \int^\infty_0 dk \frac{k\tanh(\pi k)}{2\pi} \frac{1}{\omega^2+\xi_k^2+\Delta^2}
	\end{equation}
    Regularizing the integral over $k$  to account only for allowed phonon scatterings and reducing the integral over $k$ into integral over energies with the corresponding hyperbolic density of states, this can be further simplified:
	\begin{equation}
		1=U \nu(\epsilon_\textrm{F}) T\sum_n \int_{|\xi_k|<\omega_\textrm{D}} d\xi\  \frac{1}{\omega^2+\xi^2+\Delta^2}
	\end{equation}
    Here, $\nu(\epsilon_\textrm{F})$ is the density of states at the Fermi energy, and the analytic formula for it is given in Eq.~\eqref{eq:DoS}. 
    
    Finally, the sum over Matsubara frequencies can be carried out using standard methods, which gives the final result for the gap equation:
    \begin{equation}
    \label{eqn:cont-gap-equation}
        1=U \nu(\epsilon_\textrm{F}) \int_0^{\omega_\textrm{D}} d \xi \frac{\tanh \frac{\sqrt{\xi^2+\Delta^2(T)}}{2 T}}{\sqrt{\xi^2+\Delta^2(T)}}
    \end{equation}
    which is the standard BCS gap equation with the hyperbolic DOS instead of the standard (i.e., Euclidean) one. 
    Hence, for the uniform hyperbolic space, the critical temperature can be estimated as $T_\textrm{c}\simeq 1.13\omega_{\textrm{D}} e^{-\frac{1}{U\nu (\mu)}}$.

    The derived result suggests that the phenomenology of superconductivity in the hyperbolic space is the same as that of the flat one. 
    Physically, this can be understood by noting that there is no ``geometrical defect'' present when parallel-transporting superconducting order parameters within different locally flat regions, as it is a scalar function for $s$-wave superconductivity.

\subsection{BCS theory with boundary}\label{sec:hyperspace}

Having discussed the uniform case, we continue with investigating open boundary conditions and potential boundary superconductivity in the hyperbolic plane, following in spirit our earlier discussion of Cayley trees. 
The direct approach would be to consider a hyperbolic disc with a finite radius, as such a construction provides a straightforward analog to Cayley trees with finite radius as studied in Section~\ref{sec:cayley}. 
However, due to the constraints provided by numerical approximations of Laplace-Beltrami operators in curved spaces, even the numerical solution of full non-linear self-consistent BdG equations in the presence of the boundary becomes complicated. 
Therefore, in Sec.~\ref {sec:LDOS_section}, we calculate LDOS in the presence of a hyperbolic boundary and present qualitative arguments on the nature of boundary superconductivity in the semi-infinite hyperbolic geometry. 
In the next Sec.~\ref {sec:WKB_section}, we show that while the LDOS behaves differently in continuous and discrete cases, the corresponding eigenstates are qualitatively similar. 
Then, in Sec.~\ref {sec:BCS_numerical}, we consider numerical approximations of the BdG equations in a numerically accessible thin-film geometry.

\subsubsection{LDOS in semi-infinite hyperbolic regions}\label{sec:LDOS_section}

 To have a better grasp on the qualitative understanding of mean-field superconductivity, we calculate the exact local density of states and inspect the nature of the LDOS amplification near the boundary. Noticing that, for big enough bounded hyperbolic disks (annuli), the geodesic curvature $\varkappa$ of its (outer) boundary approaches unity from above, $\varkappa\rightarrow1^+$, one concludes that the essential LDOS features can be captured by studying the idealized case of a horodisk with a horocycle boundary (i.e. region bounded by a curve with $\varkappa=1$). For completeness, we also compute LDOS in the case of a geodesic boundary ($\varkappa=0$).

The horodisc region admits a convenient choice of coordinate system. We work in the Poincaré half-plane coordinates, as a horocycle in this geometry can be conveniently represented by a horizontal ($y=\text{const.}$) line. The Laplace-Beltrami operator is
\begin{equation}\label{eq:Poincare_half_plane}
        \Delta_{\mathbb{H}^2}=y^2 \left(
        \frac{\partial^2}{\partial x^2}
        +
        \frac{\partial^2}{\partial y^2}
        \right).
    \end{equation}   
In order to find  LDOS, we solve for the Green's function of the Laplacian in Eq.~\eqref{eq:Poincare_half_plane} with appropriate boundary conditions. 
The Green's function satisfies
    \begin{equation}\label{hor_gf_problem}
        \begin{cases}
            (\omega+\Delta_{\mathbb{H}^2}) G(\omega, z, z_0 ) = y_0^2 \delta(x-x_0)
            \delta(y-y_0)           
            \\
            G(\omega, z,z_0)\big|_{z=x+i} = 0,
        \end{cases}
    \end{equation}
with $z=x+iy$. 
Here, we chose the horocycle boundary at $y=1$, and the additional $y^2$ near the Dirac delta function accounts  
for appropriate normalization due to the hyperbolic~metric.

The spectral function can be found from the Green's function by the inversion formula, which reads as
\begin{equation}
\rho(\omega,z)=-\frac{1}{2\pi i}\lim_{\epsilon\to 0^{+}}
       \bigl[ G(\omega+i\epsilon) - G(\omega-i\epsilon) \bigr].
\end{equation}
After performing tedious calculations (see App.~\ref{app:horodisk_DoS} for the derivation), one can find the LDOS in the horodisc:
       \begin{equation} \label{eq:dos_horodisk_main}
           \rho(\omega,y)
           =
           \frac{y}{\pi}
           \sum^{\infty}_{n=0}
           k_n
           I^2_{i\kappa}(k_n)
            K^2_{i\kappa}(k_n y) 
\end{equation}
where $\kappa = \sqrt{\omega - 1/4}$, $I_{i\kappa}$ and $K_{i\kappa}$ are the modified Bessel functions of imaginary index, and $k_n$ is the $n$-th root of $K_{i\kappa}(k)=0$ (the roots are ordered in descending order, $k_{n+1}<k_n$).

For comparison, we also consider LDOS in the semi-infinite hyperbolic geometry with the geodesic boundary; note that the geodesic curvature of such a boundary is $\varkappa=0$, contrary to the horocyclic boundary. 
This means that those are generally two distinct types of infinite regular boundaries in the hyperbolic space.
In particular, there is an exact mirror isometry when reflecting along a geodesic. 
This suggests that the Green's function in such geometry can be straightforwardly obtained from the image method, using only the Green's function of the infinite hyperboloid given by Eq.~\eqref{M-F_inverse}. The resulting LDOS is thus given by:
\begin{equation}\label{eq:ldos_geodesic}
    \rho(\omega,d)=(1-P_{-\frac{1}{2}+i\kappa}(\cosh 2d))\cdot \nu(\omega),
\end{equation}
where $d$ is the distance from the boundary, with $\nu(\omega)$ representing the bulk DOS of hyperboloid.

\captionsetup[subfloat]{labelformat=empty}
\begin{figure}[t!]
  \centering
  \subfloat[]{\includegraphics[width=0.48\linewidth]{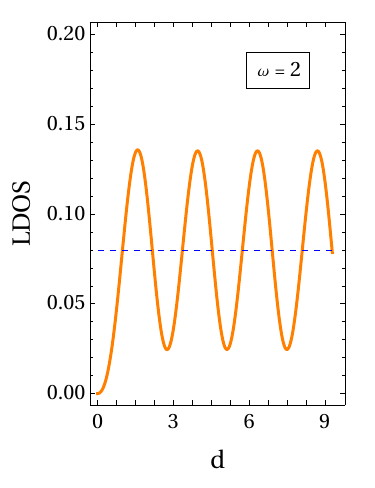}\label{fig:a}}
  \hspace{-0.01\linewidth}
  \subfloat[]{\includegraphics[width=0.48\linewidth]{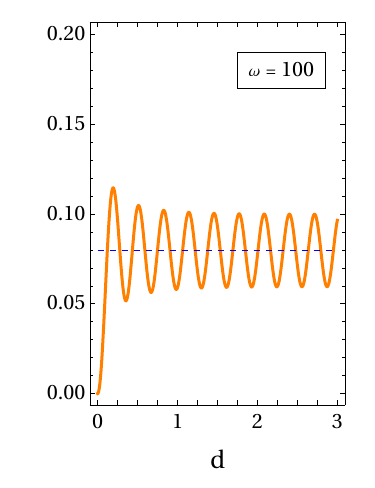}\label{fig:b}}\\[-5.5em]
  \subfloat[]{\includegraphics[width=0.48\linewidth]{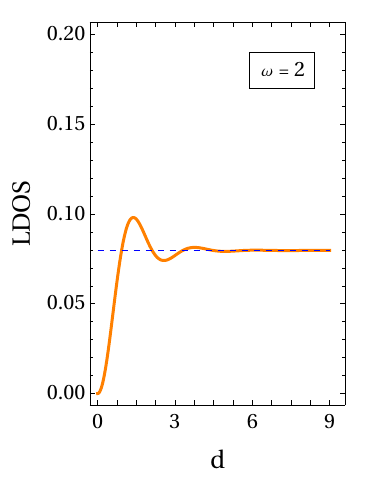}\label{fig:c}}
  \hspace{-0.01\linewidth}
  \subfloat[]{\includegraphics[width=0.48\linewidth]{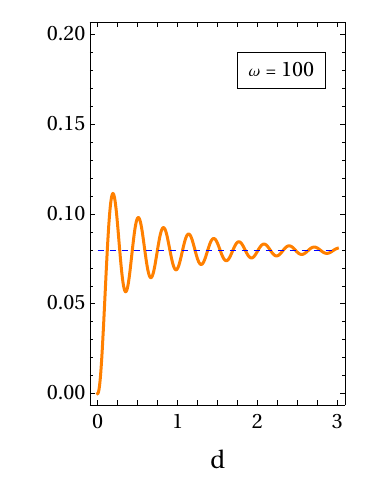}\label{fig:d}}
  \caption{Local density of states (LDOS, orange) at different values of $\omega$ as a function of the distance from the boundary $d$. The blue dashed line shows the bulk density of states (DOS) for the corresponding $\omega$. Top: LDOS in the horodisk geometry. Bottom: LDOS in the semi-infinite geometry with a geodesic boundary.}
  \label{fig:ldos_profiles}
\end{figure}
\captionsetup[subfloat]{labelformat=parens}

Typical profiles of LDOS for both geometries are shown in Fig.~\ref{fig:ldos_profiles}. One crucial distinction between horodisc (top panels of Fig.~\ref{fig:ldos_profiles}) and geodesic boundaries (bottom panels of Fig.~\ref{fig:ldos_profiles}) is that in the case of a geodesic boundary, LDOS exponentially quickly relaxes to the bulk value on the scales of curvature radius.
It is this effect which barely allows for any boundary amplification in the geodesic geometry if the ``activation length'' obeys $\kappa^{-1} \gg R$, i.e., for small $\omega$. 
In contrast, in the case of horocycle boundary, there is no such curvature damping effect, and the strongest amplification is achieved at small $\omega$. 
Particularly, LDOS oscillates around the bulk value, and at the spectral edge $\omega\rightarrow 1/4^+$, the behavior of oscillations is actually analogous to that of a flat one-dimensional half-line, i.e. oscillating between $0$ and $2\ \operatorname{DOS}_{\text{bulk}}$ \cite{Babaev:2020}. 
Let us note that in Fig.~\ref{fig:ldos_profiles}, the case of small $\omega$ (namely $\omega=2$) is shown, where this amplification is slightly less pronounced due to the fact that LDOS dips do not vanish completely as opposed to the spectral edge. The bulk density of states is restored at infinity due to the destructive interference of waves with different momenta when integrating within a small energy window. 
A few first peaks stay the same, giving rise to the amplification on the boundary $\nu = \operatorname{LDOS}(\omega,d_{\max})/\operatorname{DOS}(\omega)\approx 2$.
This hints in favor of boundary superconductivity with a higher critical temperature than in the bulk, with a relative increase comparable with the flat 1D case.

In contrast, in the case of large $\omega$ (and therefore for short ``wavelengths'' $\kappa^{-1}\ll R$), the oscillation around the bulk value is much smaller in amplitude. 
The first peak, however, doesn't disappear and approaches the universal value in the limit of large $\omega$, both in horodisc and geodesic cases (see right side of Fig.~\ref{fig:ldos_profiles}).  
The nature of the amplification can be easily understood by analyzing the asymptotic behavior of the first factor in Eq.~\eqref{eq:ldos_geodesic}; one can use the following relation, which holds for large $\kappa$:
\footnote{This relation can be established by searching for the solutions of Eq.~\eqref{eqn:H2-Laplace-Beltrami} in the form $\psi(\varrho)=\sqrt{\varrho/\sinh\varrho}\ v(\varrho)$. Corresponding equation for $v(\varrho)$ is 
$$\varrho^2v^{\prime\prime}(\varrho)+\varrho v^{\prime}(\varrho)+
\left[(\kappa\varrho)^2+\vartheta (\varrho)\right]v(\varrho)=0,$$
with $\vartheta (\varrho)=\frac{1}{4}\left(\varrho^2 \coth^2(\varrho)-1\right)-\frac{3}{4}\rho^2$. For all $\varrho$ and large $\kappa$, $\vartheta(\varrho)\ll(\kappa\varrho)^2$, the leading solution satisfies Bessel equation, thus confirming the relation as in Eq.~\eqref{eq:P_Bessel_rel}.
 }
\begin{equation}\label{eq:P_Bessel_rel}
    P_{-\frac{1}{2}+i\kappa}(\cosh x)\approx \sqrt{
    \frac{x}{\sinh{x}}
    }
    J_0(\kappa x),
\end{equation}
where $J_0(x)$ is the Bessel function of the first kind, of the zeroth order. One can see that in the limit of large $\kappa$, the first maximum of LDOS is achieved at $d\approx j_{0,1}^\prime/2\kappa$ ($j_{0,1}^\prime$ represents the first nontrivial root of $J_0^\prime(x)=0$), with the amplification factor given by $\nu \approx1-J_0(j_{0,1}^\prime)\approx 1.4028$, i.e. exactly the same as in the flat 2D semi-infinite case. 
Therefore, we expect boundary superconductivity in the continuous hyperboloid at large chemical potentials to have phenomenology analogous to that of the flat 2D case \footnote{
It is also interesting to note that, in the case of the horocycle boundary and $\kappa{\gg}1$, we numerically find that the profile of LDOS given by the rather 
complicated Eq.~\eqref{eq:dos_horodisk_main} to be very well captured by the following formula, analogous to \cref{eq:P_Bessel_rel,eq:ldos_geodesic}:
\begin{equation}
    \rho(\omega,d)\approx\left(1-\sqrt{\frac{2d}{\tanh(2d)}}J_0(2\kappa d)\right) \cdot \nu(\omega).\nonumber
\end{equation}}.

Thus, one can see that curvature (or, equivalently, chemical potential) effectively governs the crossover between 2D-like and 1D-like boundary regimes.

\subsubsection{Qualitative behavior of eigenstates}\label{sec:WKB_section}

Before proceeding with the numerical solutions of mean-field superconductivity equations in the hyperbolic geometry, we present a qualitative comparison between wavefunctions of the hyperbolic disc (annuli) and Cayley trees.

First, instead of horodisc geometry, we choose a hyperbolic annulus with the inner and outer radii denoted as $R_\textrm{in}$ and $R_\textrm{out}$. 
We also adopt this geometry later for numerical solutions of BdG equations. 
The kinetic operator $h_\textrm{ring}=-\triangle_\textrm{ring}$ on the hyperbolic annulus is the negative Laplacian operator described in Eq.~\eqref{eqn:H2-Laplace-Beltrami}:
	\begin{eqnarray}
	&\phantom{=}&h_\textrm{ring} \psi(\varrho,\phi) \nonumber \\
    &=& -\Big( \frac{1}{\sinh \varrho} \frac{\partial}{\partial \varrho} \sinh \varrho \frac{\partial}{\partial \varrho} + \frac{1}{\sinh^2 \varrho} \frac{\partial^2}{\partial \phi^2}\Big)\psi(\varrho,\phi)
	\end{eqnarray}
with the additional boundary conditions: 
\begin{equation}
\psi(R_\textrm{in})=0=\psi(R_\textrm{out}).
\end{equation} 
The eigenfunctions $\psi_{k}$ of the Laplacian $\triangle_\textrm{ring}$ are normalized according to the hyperbolic metric:
\be\label{eq:initial_norm}
\int_{\textrm{Ring}}\sqrt{g}|\psi|_k^2d\varrho d\phi=1.
\ee
 
To find eigenstates obeying the last three equations, we assume that the inner radius of the disc is much larger than unity, $R_\textrm{in}\gg 1$ (in units of the curvature radius). 
Further, we can introduce the angular momentum and consider the eigenfunctions of the form $\psi(\phi,\varrho)=e^{im\phi}\psi_m(\varrho)$.
Having done that, the kinetic operator takes a new form:
\be
\label{eq:Laplacion_ring_firstapx} h_\textrm{ring}\psi_m(\varrho)=-\Big(\frac{\partial^2}{\partial \varrho^2} +\frac{\partial}{\partial \varrho} -\frac{4m^2}{e^{2\varrho}}\Big)\psi_m(\varrho)
\ee
Here, we used that $\varrho\gg 1$ inside the annulus; therefore, we approximated $\sinh\varrho\simeq e^{\varrho}/2$ and $\coth(\varrho)\simeq 1$. Accordingly, the determinant of the metric becomes~$g\simeq e^{2\varrho}/4$. 

We can simplify the kinetic part even further by making a substitution $\psi_m(\varrho)=e^{-\frac{\varrho}{2}}f_m(\varrho)$. 
This substitution allows us to remove the first derivative:
\be\label{eq:kinetic_disc}
h_\textrm{ring}f_m(\varrho)=-\Big(\frac{\partial^2}{\partial \varrho^2}-\frac{4m^2}{e^{2\varrho}}-\frac{1}{4}\Big)f_m(\varrho).
\ee
Finally, we can introduce a new coordinate $x=R_\textrm{out}-\varrho$ and the radial momenta $\kappa_m=2me^{-R_\textrm{out}}$:
\be\label{eq:kinetic_disc_x}
h_\textrm{ring}f_{\kappa_m}(x)=-\Big(\frac{\partial^2}{\partial x^2}-e^{2x}\kappa_m^2-\frac{1}{4}\Big)f_{\kappa_m}(x).
\ee
The boundary conditions are $f_{\kappa_m}(0)=f_{\kappa_m}(d)=0$, where $d=R_\textrm{out}-R_\textrm{in}$, and the normalization condition on the eigenfunctions becomes:
\be\label{eq:rho_normalization}
\pi \int^{d}_{0}|f_{\kappa_m ,n}|^2 dx=1.
\ee
There are two reasons for adopting such a change of variables. First, it becomes clear that in the limit of large ring radius $R_\textrm{out}$, the discrete angular quantum number $m$ effectively becomes a continuous variable. Second, these coordinates are more suitable for numerical solutions, since the large exponential damping seen in Eq.~(\ref{eq:Laplacion_ring_firstapx}) has been
removed in a controlled~way.

\begin{figure}[t]
    \centering
 \includegraphics[width=\linewidth]{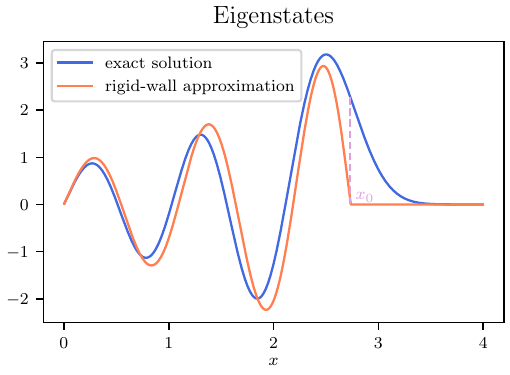} 
    \caption{The examples of the solutions of Eq.~\eqref{eq:kinetic_disc} in blue and the approximate wavefunctions described by Eq.~\eqref{eq:WKB_states} in orange. The coordinate $x=0$ corresponds to the outer boundary of the annulus. For both solutions, the angular momentum $\kappa=0.4$ and the index of the energy level is $n=5$.}
\label{fig:example_WKB_states}
\end{figure}

To highlight features of spatial the profiles of eigenstates, we apply the following rigid-wall approximation:
we assume that at the classical turning point $x_0$, defined by the condition $V(x){=}\epsilon$, there is an infinite potential barrier.
Up to a phase shift coming from the Maslov index, such an approximation will qualitatively coincide with the WKB solution, if the classically allowed region is bigger than the curvature radius, $x_0\gtrsim 1$ \cite{Maslov_book}.
In this case, one can immediately find the wave functions~$\psi_{\kappa_m}$:
\begin{gather}
\label{eq:WKB_states}
\!\psi(x)_{\kappa_m,n}=\begin{cases} \sqrt{\frac{2e^{-R_\textrm{out}}}{\pi x_0}}e^\frac{x}{2}\sin\frac{\pi n x}{x_0}, \;\, x\le x_0, \\0, \quad x> x_0,\end{cases}
\end{gather}
expressed in terms of the $x$ coordinate. 
One can see the resemblance with the nonsymmetric states in a Cayley tree as in Eq.~\eqref{eq:Cayley_tree_eigstate}.
The behavior of a typical wave-function of this form~(\ref{eq:WKB_states}) is shown in Fig.~\ref{fig:example_WKB_states}. 
For comparison, we therein also show a numerical solution of Eq.~\eqref{eq:Laplacion_ring_firstapx}. For both states, we take the same angular momentum $\kappa_m=0.4$, energy level $n=5$, and for clarity, we omit the normalization related to angular momentum. One can see that the hard wall turning point approximation demonstrates behavior similar to the exact solutions.

Nonetheless, while the profiles of wave functions in both continuous and tree geometries resemble each other, as we have seen, the boundary LDOS as a function of energy behaves in a strikingly different way. 
The boundary LDOS in Cayley is a collection of delta-functions (flat bands); at the same time, in continuous geometry, LDOS is a smooth function. 
We therefore anticipate that the boundary superconducting phenomena should be noticeably different.

\subsubsection{Numerical solution of the mean-field equations}\label{sec:BCS_numerical}

Having performed a qualitative analysis, we turn to the numerical solutions of the BdG equations. The BdG equations with the kinetic part described by Eq.~\eqref{eq:Laplacion_ring_firstapx} can be analyzed numerically, similar to the calculations performed for the Cayley trees.
We can write the BdG equations for each angular momentum $m$ separately:
\begin{gather}
\begin{pmatrix}
    h_{m}-\mu&\Delta(\varrho)\\\Delta(\varrho) &-h_{m}+\mu
    \end{pmatrix}    \begin{pmatrix}
    u_{n,m}\\v_{n,m}
    \end{pmatrix}=E_{n,m}\begin{pmatrix}
    u_{n,m}\\v_{n,m}
    \end{pmatrix},
\end{gather}
where $h_m=-\triangle_\textrm{ring}$, and $\mu$ is the chemical potential.
The solutions for the blocks are connected via the self-consistent equations for $\Delta(\varrho)$:
\be
\Delta(\varrho)=\frac{U}{2}\sum_n\sum_m u_{n,m}(\varrho)v_{n,m}(\varrho)\tanh(\frac{E_{n,m}}{2T})
\ee
where the sum is taken over all energies of BdG Hamiltonian in a chosen range. The eigenstates are normalized according to \cref{eq:initial_norm}:
\be\label{eq:BdG_norm_rho}
\int^{R_{\rm out}}_{R_{\rm in}}2\pi\sqrt{g} (|v_{n,m}|^2+| u_{n,m}|^2)d\varrho=1.
\ee

Next, we perform the same shift
of variables $x=R_\textrm{out}-\varrho$, consider large radii of the annulus $\varrho\gg1$, and introduce an exponential factor, as previously applied for studying electronic structure. Additionally, we introduce a rescaling multiplier $\frac{1}{\sqrt{\pi}}$ to simplify the subsequent normalization.
The transformation to the new wavefunctions is:
\be
\begin{pmatrix}
    u_{n,m}\\ v_{n,m}
    \end{pmatrix}=\frac{1}{\sqrt{\pi}}e^{\frac{x-R_{\rm out}}{2}}\begin{pmatrix}
    \tilde  u_{n,m}\\ \tilde  v_{n,m}
    \end{pmatrix}.
\ee
After this transformation, the kinetic operator becomes as in \cref{eq:kinetic_disc_x}, while the BdG Hamiltonian does not change its form:
\begin{gather}\label{eq:BdG_xring}
\begin{pmatrix}
    h_{m}-\mu&\Delta(x)\\\Delta(x) &-h_{m}+\mu
    \end{pmatrix}    \begin{pmatrix}
    \tilde u_{n,m}\\\tilde v_{n,m}
    \end{pmatrix}=E_{n,m}\begin{pmatrix}
    \tilde  u_{n,m}\\ \tilde v_{n,m}
    \end{pmatrix}.
\end{gather}
The normalization condition simplifies to:
\be\label{eq:BdG_norm}
\int^d_0 (|\tilde v_{n,m}|^2+|\tilde u_{n,m}|^2)dx=1.
\ee
Expressing vectors $(u,v)^{T}$ via $(\tilde u,\tilde v)^{T}$, we obtain the self-consistent equations for the BdG Hamiltonian in Eq. \eqref{eq:BdG_xring} with the normalization in Eq.~\eqref{eq:BdG_norm}:
\be
\!\Delta(x)=\frac{U}{2\pi}e^{x-R_\textrm{out}}\!\sum_n\!\sum_m \! \tilde u_{n,m}(x) \tilde v_{n,m}(x)\tanh(\frac{E_{n,m}}{2T})
\ee 
where the sum is taken over all sectors of different angular momenta, and the energies are taken within the Debye window. It is also important to note that in the limit of a large disk radius, $R_\textrm{out}\gg 1$, the summation over $m$ becomes an integral over the radial momentum $\kappa$, with the range of integration defined by the Debye window:
\be\label{eq:infiniteR_gapeq}
\Delta(x)=\frac{U}{4\pi}e^{x} \sum_n\int  \tilde u_{n,\kappa}(x) \tilde v_{n,\kappa}(x)\tanh(\frac{E_{n,\kappa}}{2T})d\kappa.
\ee
The additional prefactor $\frac{1}{2}$ occurs because changing angular quantum number $m$ by $1$ leads to the discretization step $d\kappa=2e^{-R_{\rm out}}$. 
Note that we integrate over all possible values of $\kappa$, both positive and negative.
The numerical solution of the self-consistent equations can be done by discretization of the kinetic operator and studying the BdG equations on a grid. 

There are a few additional technical details that should be discussed. 
The first question
is how to define the energy cutoff in the self-consistent equations. 
Since the density of states for a hyperbolic disc is constant at infinity, one needs to introduce regularization for the self-consistent equations. In the case of the usual BCS theory, one introduces the Debye frequency $\omega_\textrm{D}$ that bounds the energies of free electrons around the Fermi energy: $-\omega_\textrm{D}\le\xi\le\omega_\textrm{D}$. 
This scheme is difficult to apply to inhomogeneous systems, since the eigenenergies $E$ of an inhomogeneous BdG Hamiltonian depend on single-particle energies $\xi$ in a nonuniform way. 
The cutoff on the single-particle energy
can be introduced only in the vicinity of a critical point when one can solve linearized self-consistent equations. 
For this reason, we impose cutoffs $E_\textrm{min}$ and $E_\textrm{max}$ directly on the energies of the BdG Hamiltonian, such that $E_\textrm{min}\le E\le E_\textrm{max}$. 
Qualitatively, that means that the Debye
frequency becomes dependent on the order parameter. 
However, if the solution is found for some $\Delta$, it means that if one were to choose the corresponding Debye
frequency as constant, the found solution would not change. 
Therefore, this choice of regularization reproduces the critical temperature for the uniform systems with $\omega_\textrm{D}=E_\textrm{max}$, and $-\omega_\textrm{D}=E_\textrm{min}$.  
Moreover, since the summation over energies diverges only logarithmically, a different choice of regularization changes the solution of the gap equation by a multiplicative factor of the order of one. 
Finally, we should note that in the case of Cayley trees, there were no such difficulties because we integrated over all available energies $E$. 
The integral therein converges since, due to the discretization of the space into a lattice, the density of states has a finite band~width.

The numerical scheme imposes additional constraints. In particular,
we can consider only relatively small widths of the annulus, due to the discretization of the differential operator in 
Eq.~\eqref{eq:kinetic_disc_x}. We also cannot consider large chemical potentials, because the states with small $\kappa_m$ and large energies would have high oscillations, and the correct computation of large values of energies would require a larger grid. 
The last caveat is that the larger the radius of the annulus $R_\textrm{out}$, the more values of $m$ we should consider; therefore, only for low energies and relatively small $R_\textrm{out}$ we can calculate the order profile precisely. 
For larger energies, we can discretize the angular momentum $\kappa_m$ and consider this discretization as an approximation to the integral in Eq.~\eqref{eq:infiniteR_gapeq}.

 \begin{figure}[t]
    \centering
    \includegraphics[width=\linewidth]{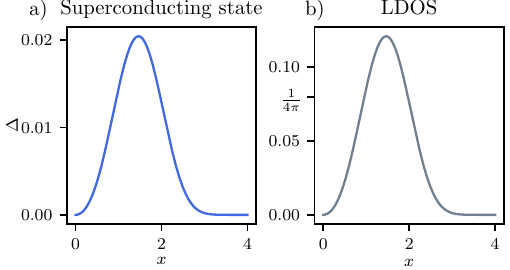} 
    \caption{
    The spatial profile of the order parameter on a hyperbolic annulus with width $d=4$. The parameters are $U=5$, $T=0.02$, $\mu=2.3$ and the minimum and maximum energies are $E_\textrm{min}=-0.2$ and $E_\textrm{max}=0.2$. The adopted discretization over angular momentum is $\textrm{d}\kappa_m=10^{-4}$. 
    The theoretically and numerically computed value of the critical temperatures in the bulk is $T^{\rm bulk}_\textrm{c}=0.0183$.}
    \label{fig:hyper_boundary_state}
\end{figure}

We finally proceed to the discussion of the numerical results. 
A prototypical profile of a superconducting state on an annulus is displayed
in Fig.~\ref{fig:hyper_boundary_state}, where we have adopted the parameters
$U=5$, $T=0.02$, $\mu=2.3$, and the energy window is given by $E_\textrm{min}=-0.2$ and $E_\textrm{max}=0.2$, which corresponds to $\omega_\textrm{D}{=}0.2$.
For a uniform system, this choice of $\mu$ gives a value of the Fermi energy at the location where the density of states is already almost constant [see Eq.~(\ref{dos_0})]; however, the results of the qualitative analysis from the previous section are not applicable for this choice of parameters, because we consider relatively small $\mu$ and annulus width $d$.
We have adopted a discretization of the angular momentum $\textrm{d}\kappa_m=10^{-4}$, which gives an exact solution on the annulus with $R_\textrm{out}\approx 10$. 
Unfortunately, we cannot directly compare the results with bulk states by considering the hyperbolic disc as we did for the Cayley trees, where we studied superconductivity on the whole tree and had direct access both to the boundary and to the center. 
In our approach to the continuous case, we focused on thin annulus geometry, where we have access only to the hyperbolic boundary. 
Nevertheless, a comparison of bulk vs.~boundary superconductivity can be achieved by comparing the annulus calculations against the calculation of 
the order parameter for the infinite space.

To find the critical temperature in the bulk with parameters corresponding to calculations made for the hyperbolic annulus, we use the results of Section~\ref {sec:uni_space}. 
In the gap equation~\eqref{eqn:cont-gap-equation}, we take $\omega_\textrm{D}=E_\textrm{max}$ and the value of the density of states being $\nu_0=\frac{1}{4\pi}$ [we take $m=\frac{1}{2}$, $\hbar=1$, $R=1$ in Eq.~\eqref{eq:DoS}], which is a good approximation for energies larger than $1$. 
The BCS answer gives us the critical temperature $T_\textrm{c}\simeq 1.13\omega_{\textrm{D}} e^{-\frac{4\pi}{U}}$. Since the values of the density of states closer to the spectral edge are lower, this critical temperature is also an upper bound for all values of the chemical potential. 
For the parameters $\omega_{\textrm{D}}=0.2$ and $U=5$, the analytical formula gives $T^{\rm bulk}_\textrm{c}=0.0183$. 
The exact numerical calculation of Eq.~\eqref{eqn:cont-gap-equation} gives a correction only for higher digits. Fig.~\ref{fig:hyper_boundary_state}(a) demonstrates that for $T{=}0.02>T^{\rm bulk}_c$ there exists a superconducting state in the hyperbolic annulus with the corresponding parameters of $\mu$, $U$, and $\omega_\textrm{D}$. 
Since the computed values of $\Delta$ are small, we can assume that $T=0.02$ is close to the critical temperature. 
Using that, we can estimate the relative increase of the critical temperature to be around $10\%$. 
Fig.~\ref{fig:hyper_boundary_state}(b) demonstrates the local density of states at $\mu=2.3$ calculated numerically using the formula:
\begin{gather}
\textrm{LDOS}(\mu, x)=\frac{1}{2\omega_\textrm{D}}\int^{\mu+\omega_\textrm{D}}_{\mu-\omega_\textrm{D}}\textrm{LDOS}(\epsilon, x)d\epsilon=\nonumber\\=\frac{e^{R_\textrm{out}}}{4\omega_\textrm{D}}\sum_n \int d\kappa |\psi_{\kappa,n}(x)|^2,\label{eqn:cond-LDOS}
\end{gather}
where $\psi_{\kappa,n}$ are eigenfunctions of Laplacian as in Eq.~\eqref{eq:Laplacion_ring_firstapx} with appropriate normalization.

The presented LDOS agrees with the mean-field calculations, since the maximum value of LDOS in the annulus is higher than the bulk value $\nu_{\rm{bulk}}=\frac{1}{4\pi}$.
While we cannot infer the existence of a separate boundary superconducting phase, the observed `thin-film' superconducting state crucially depends on the presence of boundaries. Therefore, the higher $T_\textrm{c}$ in the thin hyperbolic annulus is suggestive of a possible separate boundary superconducting phase with enhancement of boundary critical temperature similar to one-dimensional chains \cite{Babaev:2020}.

Because of the geometric complexity of the hyperbolic plane, a rigorous proof of a boundary-localized superconducting phase in the continuum remains an open problem. 
Even so, our analysis offers two complementary arguments pointing in this direction. 
First, the structure of single-particle eigenstates in hyperbolic space mirrors that of Cayley trees, thus providing the connection between the electronic structure of discrete and continuous geometries. 
Next, we performed the exact calculations for LDOS in the horodisc region and found an enhancement of the boundary LDOS.
While the enhancement is not as striking as in the case of trees, it is still larger in highly curved hyperboloids than in the flat two-dimensional spaces, and resembles the behavior of one-dimensional flat systems. 
These parallels suggest that negative curvature can stabilize boundary-localized pairing by amplifying boundary degrees of freedom. 
Finally, in annular geometries embedded in the hyperbolic space, we observe superconducting order at temperatures higher than the critical temperature of the infinite hyperbolic plane.

\section{Conclusion}

Within a mean-field framework, we have analyzed $s$-wave superconductivity in negatively curved geometries and established a coherent picture that connects discrete tree graphs and the continuous hyperbolic plane. For uniform spaces (both discrete and continuous), we derived an exact self-consistent gap equation expressed solely through the single-particle density of states, thereby recovering the standard BCS structure.

Open boundaries qualitatively change this picture. On Cayley trees, we constructed a symmetry-adapted block decomposition, which enabled self-consistent BdG calculations on trees of large radial sizes. These calculations exhibit an intermediate regime in which the superconducting order parameter is exponentially localized at the boundary while the bulk remains in the normal state, and, for sufficiently large trees, two distinct critical temperatures $T_\textrm{c}^\textrm{edge}{>}T_\textrm{c}^\textrm{bulk}$ emerge, with boundary critical temperature being significantly higher than the boundary one. 
The striking difference with flat systems, where the amplification of the boundary critical temperature is much more modest, can be possibly explained by the structure of the local density of states on the boundary: contrary to any continuous system, LDOS is not a regular function of energy but is a distribution which consists of a collection of Dirac $\delta$ peaks at rational values of momentum along the tree branches, i.e. is reminiscent of the system with many flat bands. This suggests that while the properly regularized LDOS of the tree (e.g., integrated over a finite energy support) may exhibit moderate amplifications on the boundary (analogous to continuous hyperbolic spaces), the nature of the boundary superconductivity in this case is not controlled by the usual BCS phenomenology, but by the corresponding flat-band one. 
This may qualitatively explain the significant amplification of the boundary critical temperature.

In the continuum hyperbolic plane, we investigated open boundary conditions on a hyperbolic annulus of large radius.
To understand the nature of boundary superconducting states, we have calculated the exact LDOS in the semi-infinite hyperbolic regions, and have found an enhancement of the density of states at the boundary. The found enhancement of the boundary LDOS resembles that in the flat one- and two-dimensional spaces, with the curvature (or, equivalently, chemical potential) tuning between two regimes.
To carry out qualitative analysis, we have also numerically solved mean-field equations on a hyperbolic annulus for relatively small Fermi energy and annulus width. 
In our numerical calculations, we have found a superconducting state existing for temperatures higher than the bulk critical temperature for the corresponding parameters.
While the profile of the order parameter does not demonstrate boundary localization, the appearance of the superconducting condensate is enabled by the presence of the annulus boundaries. 
Therefore, this numerical result, along with the LDOS enhancement, provides an indication of boundary superconducting states in hyperbolic geometries with a critical temperature higher than that in the bulk.

Despite the different qualitative behavior of boundary LDOS for the discrete vs.~the continuous geometry, the results suggest the existence of hyperbolic boundary superconductivity, more pronounced
than in the formerly studied Euclidean two-dimensional systems \cite{Babaev:2020,Barkman:2022, Croitoru:2020, Talkachov:2023,Hainzl:2022}.
Therefore, the discrete (Cayley-tree) and continuous (hyperbolic plane) analyses both also suggest that analogous phenomena should occur on genuine hyperbolic lattices with open boundaries, as explored further in the companion work employing Ginzburg–Landau theory~\cite{Bashmakov:2025}.

Looking ahead, several directions for future studies of hyperbolic lattices appear particularly promising in light of our results.
First, extending beyond $s$-wave to irreducible representations of the non-Eucliedan (often very high-order)
hyperbolic point groups~\cite{Chen:2023} is a natural next step. 
Due to the interplay of angular momentum of certain unconventional order parameters carrying non-zero angular momentum (e.g., $p$-wave, $d$-wave) with 
negative curvature, bulk phases of such superconductivity might be prone to spontaneously forming a vortex lattice -- even in the absence of applied magnetic
field. 
In turn, the formation of such vortices could be prevented in the boundary-only regime of such unconventional superconductivity.
To clarify the robustness of the boundary-superconducting phase (whether with $s$-wave or unconventional order parameter), it should further be  
worthwhile to incorporate fluctuations beyond the simplest mean field approximation (e.g., by virtue of DMFT on hyperbolic graphs). 
Finally, already at the single-particle level, several features of hyperbolic boundary phase, such as the enhanced boundary DOS and the Friedel oscillations, could in principle be tested in the available 
experimental platforms that can emulate
hyperbolic spaces~\cite{Kollar:2019,Lenggenhager:2021,Zhang:2022,Zhang:2023,Chen:2023b,Chen:2023c,Chen:2024,Huang:2024}.
The existing experimental platforms for emulating hyperbolic lattices do not presently allow for the realization of superconducting phases since they are either classical (electric circuits, silicon photonics)
while the quantum platform of coplanar microwave resonators is bosonic (uses photons)~\cite{Bienias:2022}. 
For this reason, a truly fermionic quantum platform capable of emulating the hyperbolic geometry should be sought to broaden an experimental window into correlated phases in curved spaces.

\section{Acknowledgments}

We thank Andrey Bagrov and Vladimir Bashmakov for valuable discussions. 
A.I.~acknowledges support from the UZH Postdoc Grant, grant No.~FK-24-104.
All authors were supported by the Starting Grant No.~211310 by the Swiss National Science Foundation.

\appendix

\section{Derivation of BCS equations for uniform lattices}\label{app:uniform_lattices}

In the derivation of BCS equations for vertex-transitive graphs, we follow the scheme applied for the continuous case in Section \ref{sec:uni_space}. 
The starting point is the same as Eq.~(\ref{Gorkov_eq}) for the continuous case, except that the LB operator $\triangle_{\mathbb{H}^2}/2m$ is replaced by the Hamiltonian $h$ defined on a graph, and the Dirac delta function $\Delta(x,x')$ in position coordinates is replaced by the Kronecker symbol $\delta_{ij}$ with site indices. Due to the vertex-transitivity of the underlying graph, we assume that the order parameter $\Delta$ is constant.

Expanding the Green's functions into their Matsubara components and the eigenfunctions of the kinetic Hamiltonian $H$, one has:
\begin{align*}
    \mathfrak{G}(\tau;i,j)&=T\sum_n e^{-i\omega_n\tau} \sum_k \psi_k(i,j) \mathfrak{G}(\omega_n,\xi_k),\\
      \mathfrak{F}(\tau;i,j)&=T\sum_n e^{-i\omega_n\tau} \sum_k \psi_k (i,j) \mathfrak{F}(\omega_n,\xi_k).
\end{align*}
In these components, the equations of motion become
 \begin{equation} 
        \begin{array}{rcc}
        (i\omega - \xi_k) \mathfrak{G}_k(\omega) + \Delta \mathfrak{F}_k(\omega) & = & 1 
        \\
        (i\omega + \xi_k) \mathfrak{F}_k(\omega) + \Delta \mathfrak{G}_k(\omega) & = & 0,
        \end{array} 
    \end{equation}
    which provides a solution
	\begin{align}
		\mathfrak{G}_k(\omega)&=-\frac{i\omega+\xi_k}{\omega^2+\xi_k^2+\Delta^2}\\
		\mathfrak{F}_k(\omega)&=\frac{\Delta}{\omega^2+\xi_k^2+\Delta^2}
	\end{align}
The value of the superconducting gap can now be found from the self-consistency condition $\Delta=U\mathfrak{F}(0+;0,0)$, which gives
\begin{eqnarray}
    \Delta &=& U T\sum_n \sum_k \frac{\Delta}{\omega_n^2+\xi_k^2+\Delta^2} \nonumber \\
    &=& UT\sum_n\int^{\infty}_{-\infty}\frac{\Delta \varrho(\lambda)d\lambda}{\omega^2_n+(\lambda-\mu)^2+\Delta^2}.
\end{eqnarray}
Taking the summation over Matsubara frequencies \cite{Abrikosov:107441, Altland_book}, we obtain the gap equation shown in Eq.~\eqref{eq:gapselfconst_uniform} of the main text.

	\section{ Mehler-Fock transform and density of states of hyperbolic Fermi gas}\label{app:MF_sec} 

        \begin{figure}
    \centering
    \includegraphics[width=\linewidth]{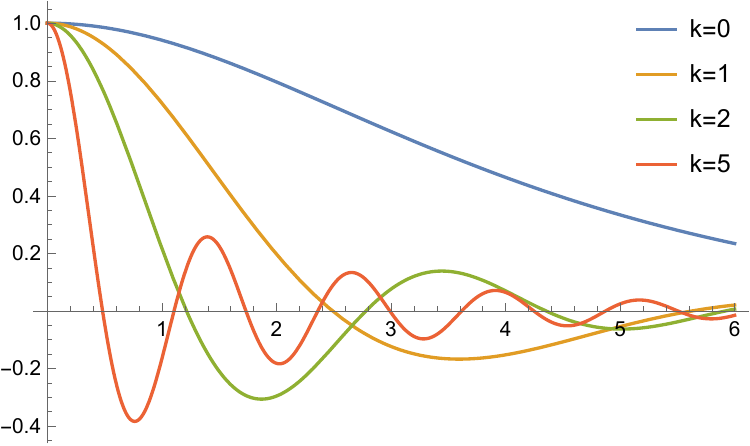} 
    \caption{The profiles of conical functions $P_{-\frac{1}{2}+ik}(\cosh \varrho)$ for several values of $k$.}
    \label{fig:Meller_functions}
\end{figure}
    
	To stick to the scope of the main text, we derive DOS from thermal Green's functions. 
    The free thermal Green's function (in Matsubara space) satisfies the equation, which can be seen as Eq.~(\ref{eqn:cont-greeens}) in the normal state, i.e., when $\Delta= 0$:
	\begin{equation}\label{FreeGF}
		\left\{i\omega_n + \frac{\triangle_{\mathbb{H}^2}}{2m}\right\}\mathfrak{G}_0(i\omega_n,\varrho)=\frac{\delta(\varrho)}{2\pi\sinh(\varrho)}=\frac{1}{2\pi}\delta(\cosh\varrho-1).
	\end{equation}
	One can see that the function on the right side satisfies the condition of a hyperbolic delta function localized at $\varrho =0$, i.e., 
    \be
    \int_{\mathbb{H}^2} \textrm{d}\varrho \textrm{d}\phi \sqrt{g} \frac{\delta(\varrho)}{2\pi\sinh(\varrho)} f(\varrho) = f(0).
    \ee
    The Mehler-Fock \cite{NIST:DLMF} transform is defined as
	\begin{equation}\label{M-F}
		\mathfrak{G}(k) = k\tanh(\pi k) \int_1^\infty P_{-\frac{1}{2}+ik}(x) \mathfrak{G}(x) dx \quad (0\leq k \leq \infty),
	\end{equation}
	and the inverse Mehler-Fock transform is
	\begin{equation}\label{M-F_inverse}
		\mathfrak{G}(x) = \int_0^\infty P_{-\frac{1}{2}+ik}(x) \mathfrak{G}(k) dk\quad (1\leq x \leq \infty),
	\end{equation}
	In the previous two equations, $P_{-\frac{1}{2}+ik}(\cosh \varrho)$ are rotationally invariant eigenfunctions of the Laplace-Beltrami operator with eigenvalues $-{k}^2{-}1/4{\equiv} {-}2m\epsilon_k$. 
    Examples of the conical (Mehler) functions $P_{-\frac{1}{2}+ik}(\cosh \varrho)$ for various values of $k$ are shown~in~Fig.~\ref{fig:Meller_functions}.

    It is convenient to use Mehler expansion to solve Eq.~(\ref{FreeGF}).
    Indeed, in the Matsubara-Mehler representation the Green's function is $\mathfrak{G}_0(i\omega_n,\varrho)=\int_0^\infty P_{-\frac{1}{2}+ik}(\cosh\varrho) \mathfrak{G}_0(i\omega_n,k) dk$, and the equation of motion reads
	\begin{equation}
		\left\{i\omega_n - {\epsilon_k}\right\}\mathfrak{G}_0(i\omega_n,k)=k\tanh(\pi k)/2\pi
	\end{equation}
	where on the right-hand side we used the expansion
    \be
    \frac{\delta(\cosh {\varrho} - 1)}{2\pi}=\frac{1}{2\pi}\!\int_0^\infty \!\!\!\!\!\! P_{-\frac{1}{2}+ik}(\cosh\varrho) k\tanh(\pi k)dk
    \ee
    which is the Mehler-Fock expansion of the delta-function in hyperbolic space.
    Thus, we find
	\begin{equation}
	\mathfrak{G}_0(i\omega_n,k)=\frac{1}{2\pi}\frac{k\tanh(\pi k)}{i\omega_n - \epsilon_k},
	\end{equation}
	which corresponds to Eq.~(\ref{eqn:cont-thermal-G-Green}) after setting $\Delta=0$. By analytic continuation, we can obtain 
    the retarded Green's function $G^\textrm{R}(\omega,k)=k\tanh(\pi k)/(\omega - \epsilon_k+i\delta)$. Density of states is $\nu(\omega)=-\frac{1}{\pi}\mathfrak{Im}G^\textrm{R}(\omega,\varrho\rightarrow0)=-\frac{1}{\pi}\int dk \mathfrak{Im}G^\textrm{R}(\omega,k)$. This gives
	
	\begin{equation}\label{dos_0}
		\nu(\epsilon)=\int_0^\infty \frac{dk}{2\pi} k\tanh(\pi k R)\delta\left(\epsilon -\frac{ \hbar^2k^2+\hbar^2/4R^2}{2m}\right),
	\end{equation}
	where, to obtain the general form of the equation, we reconstructed the curvature radius $R$ and the Planck's constant $\hbar$ from dimensional considerations.

    After integration, Eq.~(\ref{dos_0}) gives
    \begin{equation}\label{eq:DoS}
        \nu(\epsilon)=\frac{m}{2\pi\hbar^2} \tanh(\pi R\sqrt{2m(\epsilon-\epsilon_0)/\hbar^2}), \quad \epsilon\geq \epsilon_0
    \end{equation}
    with $\epsilon_0\equiv\hbar^2/8mR^2$ being the ``zero-point'' energy of the Laplace-Beltrami operator in hyperbolic plane. 
    Note that in the vicinity of the band edge, 
    the scaling law of DOS is $\nu(\epsilon)\propto (\epsilon-\epsilon_0)^{1/2}$. Interestingly, this scaling law is different from Euclidean systems in two dimensions, and it instead matches Euclidean systems in three spatial dimensions.

    \section{Local density of states in the horodisk}\label{app:horodisk_DoS} 

    In this appendix, we derive Green's functions and local density of states in the horodisc. We start with Eqs.~\eqref{eq:Poincare_half_plane} and \eqref{hor_gf_problem} from the main text, which assume the half-plane representation of the hyperbolic plane.
    Translation symmetry of the problem along the $x$-direction suggests the Fourier representation
    \be
    G(\omega,z,z_0)=\int_{-\infty}^{\infty}\frac{dk}{2\pi}\ e^{-ik(x-x_0)}y_0^2G_k(\omega, y,y_0).
    \ee
    In terms of Fourier components, the problem reduces to the one-dimensional one:
    \begin{equation}
       \left[ 
            y^2 \frac{\partial^2}{\partial y^2}
            -k^2y^2
            +z
       \right]G_k=\delta(y-y_0).
    \end{equation}
    By introducing $g=(y_0/y)^{\frac{1}{2}}G_k$, Eq.~(\ref{hor_gf_problem})
    can be further
    reduced to:
    \begin{equation}\label{eq:gf_hor}
        \begin{cases}
            y^2g^{\prime\prime}
            +
            yg^{\prime}
            +
            \left[
                (\omega-1/4)
                -
                k^2 y^2
            \right] g
            = 
            \delta(y-y_0)           
            \\
            g(1,y_0)=0.
        \end{cases}
    \end{equation}
The solution of this 1D Green's function problem has a general form:
\begin{equation}
    g=\begin{cases}
        \frac{u_<(y)u_>(y_0)}{y_0^2\mathcal{W}(y_0)},\ 1\leq y < y_0\\
        \frac{u_<(y_0)u_>(y)}{y_0^2\mathcal{W}(y_0)},\ y_0 < y < \infty ,
    \end{cases}
\end{equation}
where $u_<(y)$ and $u_>(y)$ are linearly independent solutions of the corresponding linear homogeneous equation, subject to boundary conditions $u_<(1)=0$ and $u_>(\infty)=0$, and where $\mathcal{W}[u_<,u_>]=u_<(y)u^\prime_>(y)-u_<^\prime(y)u_>(y)$ is their Wronskian determinant.

\begin{figure}[!t]
    \centering
    \includegraphics[width=\linewidth, trim={0cm 0cm 0cm 0.9cm}, clip]{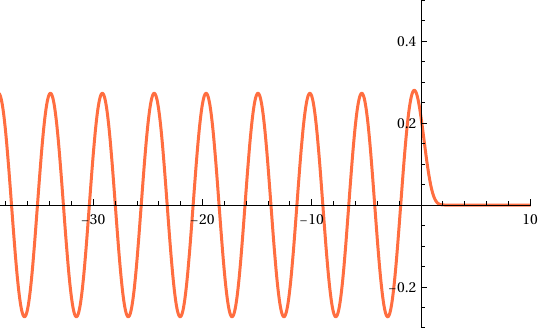} 
    \caption{The typical profile of $K_{i\kappa}(e^d)$; here $\kappa = \sqrt{2-1/4}$.}
    \label{fig:Modified_Bessel_functions}
\end{figure}

One can check that the following pair of solutions of the homogeneous equation corresponding to Eq.~(\ref{eq:gf_hor}) satisfies the required boundary conditions:
\begin{align}
    u_<(y)& = 
            I_{\sqrt{\frac{1}{4}-\omega}}(|k|y)
            -
            \frac{
                I_{\sqrt{\frac{1}{4}-\omega}}(|k|)
            }{
                K_{\sqrt{\frac{1}{4}-\omega}}(|k|)
            }
            K_{\sqrt{\frac{1}{4}-\omega}}(|k|y)
         \\
    u_>(y)&=K_{\sqrt{\frac{1}{4}-\omega}}(|k|y),
\end{align}
where $I_\nu(y),K_\nu(y)$ denote the modified Bessel functions, and the Wronskian determinant is~\cite{NIST:DLMF}
\begin{equation}
\mathcal{W}[u_<,u_>]=\mathcal{W}[I_{\sqrt{\frac{1}{4}-\omega}}(|k|y),K_{\sqrt{\frac{1}{4}-\omega}}(|k|y)]=-1/y.
\end{equation}
In the context of DOS calculations, our primary interest lies in the diagonal components of the GF, which is finally given by
\begin{widetext}
\begin{equation} \label{eq:hor_gf_final}
        G(\omega,z,z)=-\frac{y}{\pi}\int_{0}^{\infty}{dk}\ \  
        K_{\sqrt{\frac{1}{4}-\omega}}(ky)
        \left(
                 I_{\sqrt{\frac{1}{4}-\omega}}(ky)
                    -
                    \frac{
                        I_{\sqrt{\frac{1}{4}-\omega}}(k)
                    }{
                        K_{\sqrt{\frac{1}{4}-\omega}}(k)
                    }
                    K_{\sqrt{\frac{1}{4}-\omega}}(ky)
            \right).    
\end{equation}
\end{widetext}
The spectral function can be obtained by the inversion formula
\be
\rho(\omega,z)=-\frac{1}{2\pi i}\operatorname{Disc} G(\omega,z,z),
\ee
where 
\be\operatorname{Disc} f(\omega)
    := \lim_{\epsilon\to 0^{+}}
       \bigl[ f(\omega+i\epsilon) - f(\omega-i\epsilon) \bigr].
       \ee
To apply this formula, we use the following properties of modified Bessel functions \cite{NIST:DLMF}: 
\begin{subequations}
\label{eq:connection_formula}
\begin{align}
           &K_{-\nu}(y) = K_{\nu}(y) \\
           &I_{-\nu}(y) = I_{\nu}(y)+(2/\pi) \sin({\nu \pi} )  K_{\nu}(y).
\end{align}
\end{subequations}
We focus on the case $\omega >1/4$, and introduce $\sqrt{\omega-1/4}=\kappa$. 

We now calculate the $\operatorname{Disc}$ value of the second term in Eq.~(\ref{eq:hor_gf_final}), finding
\begin{gather} \label{eq:disc_ini}
        \operatorname{Disc}
        \left(
                    \frac{
                        I_{\sqrt{\frac{1}{4}-\omega}}(k)
                    }{
                        K_{\sqrt{\frac{1}{4}-\omega}}(k)
                    }
                    K_{\sqrt{\frac{1}{4}-\omega}}(ky)
            \right)
            =\\=
            \lim_{\delta\rightarrow0^+} 
            \left(
                \frac{
                        I_{-i\kappa}(k)
                    }{
                        K_{-i\kappa+\delta}(k)
                    }
                    K_{-i\kappa
                }(ky)
                -
                 \frac{
                        I_{i\kappa}(k)
                    }{
                        K_{i\kappa+\delta}(k)
                    }
                    K_{i\kappa
                }(ky)
                \right).\nonumber
\end{gather}
In the last expression, the infinitesimal $\delta$ was kept to control the poles of $ 1/K_{i\kappa}(k)$ (see Fig.~\ref{fig:Modified_Bessel_functions}).

As such
\begin{align}
       \frac{1}{ K_{i\kappa+\delta}(k)}
       &\approx 
       \frac{1}{ K_{i\kappa}(k)-i\delta \frac{\partial K_{i\kappa}(k)}{\partial \kappa}}
       = \\
       =\ &i\pi 
       \operatorname{sign}\left(\frac{\partial K_{i\kappa}(k)}{\partial \kappa}\right)
       \delta( K_{i\kappa}(k))
       +
       \mathcal{P}
       \left(
            \frac{1}{ K_{i\kappa}(k)}
       \right)\nonumber
\end{align}
where $\mathcal{P}$ stands for the principal value.
For brevity, we further denote $s_{\kappa,k}= \operatorname{sign}\left(\frac{\partial K_{i\kappa}(k)}{\partial \kappa}\right)$, and with $\delta( K_{i\kappa}(k))$ we mean
the Dirac delta function of $K_{i\kappa}(k))$. 
Using Eqs.~(\ref{eq:connection_formula}), the first term in Eq.~\ref{eq:disc_ini}
can be reduced to
\begin{align} \label{eq:disc} 
        \operatorname{Disc}&
        \left(
                    \frac{
                        I_{\sqrt{\frac{1}{4}-\omega}}(k)
                    }{
                        K_{\sqrt{\frac{1}{4}-\omega}}(k)
                    }
                    K_{\sqrt{\frac{1}{4}-\omega}}(ky)
            \right)
            =\\
            =\ &K_{i\kappa}(ky)
            \left(
                   \frac{2}{\pi}
                   \sinh(\pi \kappa)
                   -2i\pi s_{\kappa,k}
                   I_{i\kappa}(k)
                   \delta\left( K_{i\kappa}(k) \right)
           \right),\nonumber
\end{align}
where we used that $K_{i\kappa}(k)\delta\left( K_{i\kappa}(k) \right)=0$ and that $K_{i\kappa}(k)\mathcal{P}
       \left(
            \frac{1}{ K_{i\kappa}(k)}
       \right)=1$. 

The $\operatorname{Disc}$ value of the first term inside the parentheses in
Eq.~(\ref{eq:hor_gf_final}) can be also
computed using the connection formulae (\ref{eq:connection_formula}); as such, the spectral function is:
\begin{align} \nonumber
           \rho(\omega,y)&=
           \frac{y}{\pi}
           \int_{0}^{\infty}{dk}\ 
           s_{\kappa,k}
           \delta\left( K_{i\kappa}(k) \right)
            I_{i\kappa}(k)
            K^2_{i\kappa}(ky)  
               \\   \label{eq:spect_func}
            =&         
           \frac{y}{\pi}
           \sum^{\infty}_{n=0}
           \operatorname{sign}\left(\frac{\partial K_{i\kappa}(k_n)}{\partial \kappa}\right)
           \frac{I_{i\kappa}(k_n)}{|K_{i\kappa}^\prime(k_n)|}
            K^2_{i\kappa}(k_n y), 
\end{align}
where $k_n$ is the $n$-th root of $K_{i\kappa}(k)=0$ (the roots are ordered in descending order, $k_{n+1}<k_n$).
Eq.~(\ref{eq:spect_func}) can be further simplified using the following identities. 
First, using the fact that the Wronskian determinant $\mathcal{W}[I_{i\kappa}(k), K_{i\kappa}(k)]=-\frac{1}{k}$, it follows for $k=k_n$ that
\begin{equation}
\label{eq:KI_relation}
  \frac{1}{K^\prime_{i\kappa}(k_n)} = - k_n I_{i\kappa}(k_n).
\end{equation}
Moreover, it can be shown that the solutions to the Sturm-Liouville problem corresponding to the homogeneous part of Eq.~(\ref{eq:gf_hor}), namely $K_{i\kappa}(ky)$, satisfy the  
normalisation condition
\begin{equation}
		\int_{1}^{\infty} \frac{dy}{y} K^2_{i \kappa} (ky) 
		=
		-\frac{k}{2\kappa} 
		K^{\prime}_{i \kappa} (k)
		 \left[
		 	\frac{\partial K_{i\kappa}(k)}{\partial \kappa}
		 \right],
\end{equation}
 for all $\kappa, k>0$ satisfying the quantization condition $ K_{i \kappa} (|k|) = 0$.
The last relation allows one to link the sign of the derivative with respect to the order of the modified Bessel function to the sign of its derivative with respect to the argument at the same point; specifically:
\begin{equation}
    \operatorname{sign}\left(\frac{\partial K_{i\kappa}(k_n)}{\partial \kappa}\right)
    =
    -
    \operatorname{sign}\left( K^\prime_{i\kappa}(k_n)\right)
    =
    - (-1)^n
\end{equation}
Therefore, we obtain:
\begin{equation}
           \rho(\omega,y)
           =
           -
           \frac{y}{\pi}
           \sum^{\infty}_{n=0}
           (-1)^n
           k_n
           \ I_{i\kappa}(k_n)
           \left|I_{i\kappa}(k_n)\right|
            K^2_{i\kappa}(k_n y).
\end{equation}
While this expression may seem complex-valued due to the presence of generally complex $I_{i\kappa}(k)$, note that taking the imaginary part of the Wronskian and remembering that $K_{i\kappa}(k)$ is a real-valued function, one can deduce that $\Im \left(I_{i\kappa}(k)\right)$ is linearly dependent from  $K_{i\kappa}(k)$ and thus has zeros at the same $k$ as $K_{i\kappa}(k)$: specifically,  $\Im \left[I_{i\kappa}(k_n)\right]=0$ and $I_{i\kappa}(k_n)=\Re[I_{i\kappa}(k_n)]$.
Furthermore, 
Eq.~(\ref{eq:KI_relation}) implies that $\operatorname{sign}\left( I_{i\kappa}(k_n)\right)=-\operatorname{sign}\left( K^\prime_{i\kappa}(k_n)\right)$. 
We thus finally have:
\begin{equation}
           \rho(\omega,y)
           =
           \frac{y}{\pi}
           \sum^{\infty}_{n=0}
           k_n
           I^2_{i\kappa}(k_n)
            K^2_{i\kappa}(k_n y),
\end{equation}
which corresponds to Eq.~(\ref{eq:dos_horodisk_main}) in the main text.

\bibliography{bib}

\end{document}